\documentclass{aa}
 \usepackage[varg]{txfonts}
 \usepackage[switch]{lineno}
 \usepackage{graphicx, epsfig, fancyhdr, rotating, amsmath, epsf, txfonts, natbib, epstopdf, multirow}
\usepackage{subfigure, tabularx, appendix, dcolumn, threeparttable, longtable, makecell}
\usepackage{psfig}
\usepackage[colorlinks=true,allcolors=blue]{hyperref}
\usepackage[normalem]{ulem}
\usepackage{hyperref}
\newcount\inclfigs \inclfigs=1
\usepackage{longtable}
\usepackage{multirow}
\usepackage{array}
\usepackage[dvipsnames]{xcolor}   

\def\arcsec{\hbox{$^{\prime\prime}$}}

\begin{document}

\title{Coronal magnetic field
and emission properties of small-scale bright and faint loops in the quiet Sun}

\author{Maria~S. Madjarska\inst{1, 2}, Thomas Wiegelmann\inst{1}, Pascal D\'emoulin\inst{3}, Klaus Galsgaard\inst{4}}

\offprints{mmadjarska@space.bas.bg}
\institute{
Max Planck Institute for Solar System Research, Justus-von-Liebig-Weg 3, 37077, G\"ottingen, Germany
\and
Space Research and Technology Institute, Bulgarian Academy of Sciences, Acad. Georgy Bonchev Str., Bl. 1, 1113, Sofia, Bulgaria
\and
LESIA, Observatoire de Paris, Universit\'e PSL, CNRS, Sorbonne Universit\'e, Université Paris Cit\'e, 5 place Jules Janssen, 92195 Meudon, France; Laboratoire Cogitamus, F-75005 Paris, France
\and
School of Mathematics and Statistics, University of St Andrews, North Haugh, St Andrews, KY16 9SS, Scotland, UK}

\date{Received date, accepted date}

\abstract
{The present study provides statistical information on the coronal magnetic field and intensity properties of small-scale bright and faint loops in the quiet Sun.}
{We aim to quantitatively investigate the morphological and topological properties of the coronal magnetic field in bright and faint small-scale loops, with the former known as coronal bright points (CBPs). }
{We analyse 126 small-scale loops of all sizes using quasi-temporal imaging and line-of-sight magnetic field observations. These observations are taken by the Atmospheric Imaging Assembly (AIA) in the  Fe~{\sc xii} 193~\AA\ channel and the Helioseismic Magnetic Imager (HMI) on board the Solar Dynamics Observatory. We employ a recently developed automatic tool that uses a linear magneto-hydro-static (LMHS) model to compute the magnetic field in the solar atmosphere and automatically match individual magnetic field lines with small-scale loops.}
{For most of the loops, we automatically obtain an excellent agreement of
the magnetic field lines from the LMHS model and the loops seen in the AIA 193~\AA\ channel. One stand-out result is that  the magnetic field is non-potential.  We obtain the typical ranges of loop heights, lengths, intensities, mean magnetic field strength along the loops and at loop tops, and magnetic field strength at loop footpoints. We investigate the relationship between all those parameters. We find that loops below the classic chromospheric height of 1.5~Mm are flatter suggesting that non-magnetic forces (one of which is the plasma pressure) play an important role below this height.  We find a strong correlation (Pearson coefficient of 0.9) between loop heights and lengths. An anticorrelation is found between the magnetic field strength at 
loop tops and loop heights and lengths. The average intensity along the loops correlates stronger with the average magnetic field along the loops than with the field strength at loop tops.}
{The latter correlation indicates
that the energy release in the loops is more likely linked to the average
magnetic field along the loops than the field strength on the loop tops. In other
words, the energy is probably released all along the loops, but not just at the
loop top. This result is consistent with the recent benchmarking radiative 3D MHD model of N\'obrega-Siberio et al.}

\keywords{Sun: chromosphere -- Sun: corona -- Sun: activity -- Sun: magnetic fields -- Methods: observational, data analysis}
\authorrunning{Madjarska et al.}

\maketitle
\section{Introduction}
\label{intro}

Coronal loops at all scales are the solar phenomena that dominate the solar corona when observed in extreme ultraviolet (EUV) emission from plasma heated to a million degrees. Loops in active regions (ARs) have been very intensively studied, and to a large extent, their morphological, magnetic, and plasma properties are well known \citep[][ and references therein]{2014LRSP...11....4R}. Bright small-scale loops in the quiet Sun, known as Coronal Bright Points (CBPs), were also studied in some details  \citep[for review see][]{2019LRSP...16....2M} although spatial and time resolution limitations have restricted certain investigations. Outside small- and large-scale loop regions, that is CBPs, and within ARs, the solar corona is filled with diffuse emission. \cite{2023ApJ...942....2T}  suggested that most of the AR heating to  high temperature is transient, while the background emission in ARs results from small steady
heating. \citet{Klimchuk_etal:1995} put forward the idea that the diffused quiet Sun emission component consists of fainter and less clearly distinguishable loops. 
\citet{Madjarska_etal:2023} demonstrated that this interpretation is indeed highly probable after deriving the properties of seven faint loops.
The morphology and the physical properties of CBP individual loops have been given less attention. One reason is that before the TRansition Region and Coronal Explorer (TRACE), the Solar Dynamics Observatory (SDO), and the IRIS space missions, the limited spatial resolution of the existing instruments did not permit resolving individual loops.
We will review hereafter some of the known properties of CBPs and their coronal magnetic structure. 

The global sizes of CBPs have been typically determined by the diameter of their on-disk projected areas, assuming a circular shape. This approach was introduced at the time of the CBP discovery in 1969 \citep{Vaiana_etal:1973} as they appeared in the limited-resolution of the first X-ray images as compact circle-like areas of enhanced emission with a diameter ranging from 20\arcsec\ to 30\arcsec\ \citep{Golub_etal:1977}. Although several studies also reported on the size of CBPs, all of them have used the same approach and found similar sizes \cite[for more details see section~3.4 in][]{2019LRSP...16....2M}. 

The heights of CBP loops, observed in extreme ultraviolet (EUV), i.e. mainly at coronal temperatures, have been estimated to range from 5 to 10~Mm with an average height of 6.5~Mm. Various methods have been employed and details can be found in section~3.5 of \citet{2019LRSP...16....2M}. To determine the heights of CBP loops in various temperatures, only CBPs nearby ARs have been studied as they are typically larger. Loops in emission from spectral lines with high formation temperatures were found to overlay
cooler once. The CBP heights in chromospheric and transition region temperatures were estimated to extend to heights of 3~Mm (with 50\%\ errors of this estimation). 

The coronal magnetic topology of CBPs is crucial in modelling
these phenomena. It was first derived by \citet{Parnell_etall:1994} and \citet{Mandrini:1996}. \citet{Perez_Suarez_etall:2008} obtained the 3D structure
of a CBP observed with several instruments on board TRACE, Solar Heliospheric Observatory (SoHO), and Hinode employing the Mpole code of \citet{Longcope:1996}. From the visual comparison of the appearance of the CBP in enhanced Hinode's X-ray Telescope (XRT) images and the extrapolated magnetic 
field (based on SoHO's Michelson Doppler Imager (MDI) magnetograms), \citet{Perez_Suarez_etall:2008} concluded that a large fraction of the magnetic field that builds up the skeleton of the CBP is close to potential.
The loop lengths were reported in two studies. \citet{Mondal_etall:2023} employed a potential field model for a single CBP and from all computed field lines they estimated that lengths are in the range from a few megameters to up to 80~Mm, with a distribution that peaks at 30~Mm. \citet{Gao_etall:2022} also report the lengths of CBP loops obtained as $L$ = $\pi{D}$/2, where $D$ is the distance of the two footpoints of a loop. The authors find a length range of 14--42~Mm, with an average of 23.5~Mm. 

The coronal magnetic properties of small-scale loops in the quiet Sun and coronal holes have been investigated, to the best of our knowledge, in a relatively small number of studies. \citet{wiegelmann_etall:2010} used data from {\rm Sunrise}/IMaX \citep{martinez_pillet_etall:2011}. The study performed potential field extrapolations (linear force-free extrapolation was also tested). The force-free extrapolation was justified by the study of \citet{martinez_gonzalez_etall:2010} which noted that the loop topology appears 
potential as the magnetic fields at the footpoints become almost vertical while the loop crosses the minimum
temperature region. \citet{wiegelmann_etall:2010} investigated all magnetic loops that connect photospheric fluxes and found an average loop height of 1.24$\pm$2.45~Mm with the magnetic field strength in the two footpoints of loops showing large differences.

Generally, there has been no attempt to estimate the length and heights of small-scale loops by directly matching individual loops with model-produced magnetic field lines. Although high-resolution images have existed since the time of the TRACE mission, more attention has been paid to AR loops, and attempts were made to find a methodology which can be used to extract the coronal magnetic properties of loops \citep{Carcedo_etal:2003}. 

The main goal of the present study is to provide statistical information about the magnetic and morphological properties of small-scale loops including their length, height, and magnetic field along the loops.  
The magnetic field observations are taken by the Helioseismic Magnetic Imager \citep[HMI,][]{Scherrer_etal:2012} and the imaging data by the Atmospheric Imaging Assembly \citet[AIA][]{Lemen_etal:2012} in the Fe~\textsc{xii} channel  on board SDO \citep{Pesnell_etal:2012}. These observations are described in full detail in \citet[][hereafter Paper~I]{Madjarska_etal:2023}. The enumeration of the loop systems is identical to those in Paper~I. 

The paper is organized as follows: In section~\ref{obs} we briefly describe the data. Full details on the data are given in Paper~I. Section~\ref{method} describes in short the methodology for computing linear magneto-hydro-static (LMHS) equilibria. 
Section~\ref{res} presents the results and discussion including presentation and discussion on the obtained LMHS parameters (Section~\ref{lmhspar}), the loop physical parameters (Section~\ref{looppar}), and their relationship (Section~\ref{relation}). Section~\ref{sum_concl} gives a summary and the conclusions.

\section{Observational data}
\label{obs}

The observations used in this study were described in detail in Paper~I. Here, we summarise the main details of these data. We utilised imaging data from AIA \citep{Lemen_etal:2012} onboard SDO \citep{Pesnell_etal:2012} taken in the 193~\AA\ channel (hereafter AIA 193) and HMI/SDO line-of-sight magnetograms \citep{Scherrer_etal:2012}. The data were collected over a period of 48 hours starting on 2019 September 15 at 00:00 UT. The AIA 193 data have a 12-second cadence and 0.6\arcsec\ $\times$ 0.6\arcsec\ pixel size. To increase the signal-to-noise ratio, we binned every three consecutive AIA 193 images. We used HMI line-of-sight magnetograms taken at a 45-second cadence. The HMI data originally have a 0.5\arcsec\ $\times$ 0.5\arcsec\ pixel size but were rescaled to the AIA pixel size of 0.6\arcsec\ using the hmi\_prep.pro procedure. To increase the signal-to-noise ratio, we binned eight consecutive magnetograms. All images were derotated to 00:00 UT on September 16, 2019. The final cadence of the images is approximately 6 minutes. The events selected for this study are taken from a square field-of-view that covers -400\arcsec\ to 400\arcsec\ from the disk centre (or 1334 $\times$ 1334  px$^2$) (see fig.~1 in Paper~I). The enumeration of the CBPs in the present paper is the same as in Paper~I, allowing the reader to identify each CBP location in Fig. 1 of Paper~I. The bright and faint loop systems were visually selected. The selection approach is described in full detail in Section 3 of Paper~I.

\begin{figure}[t!]
    \centering
    \includegraphics[width=.5\textwidth]{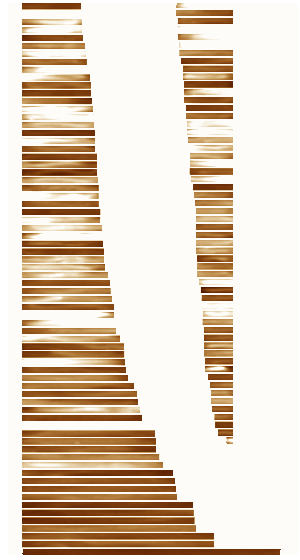}
    \caption{All 126 examples of the uncurled loops showing their relative lengths and intensity profiles. The colour level shows the AIA~193 intensity on a linear scale. 
    The loops are arranged by the loop length.}
    \label{fig1}
\end{figure}

\section{Methodology: MHS equilibria}
\label{method}
The present study employs a recently developed automatic algorithm to compute 
LMHS equilibria and match obtained field lines with features with enhanced emission in the SDO/AIA~193 channel \citep[][hereafter Paper~II]{Wiegelmann_Madjarska:2023}.   In the following, we will briefly describe the algorithm.  
The code uses line-of-sight HMI magnetograms as the photospheric boundary condition. While the solar corona is assumed to be force-free due to
the low plasma $\beta$, which is not the case in the lower solar atmosphere.
In the photosphere and chromosphere, the Lorentz force remains finite and
needs to be compensated in the magneto-hydro-static (MHS) approximation by
the plasma pressure gradient force and the gravity force. The MHS equations are:
\begin{eqnarray}
{\bf j}\times{\bf B} & = &  \nabla p +\rho\, \nabla \Psi,
\label{forcebal}\\
\nabla \times {\bf B } & = &  \mu_0\, {\bf j} , \label{ampere} \\
\nabla\cdot{\bf B}  & = &  0,   \label{solenoidal}
\end{eqnarray}
where ${\bf B}$ is the magnetic field,
${\bf j}$ the electric current density, $\mu_0$ the permeability of free space,
$p$ the plasma pressure,
$\rho$ the mass density, and $\Psi$ the gravitational potential.
\citet{2022ScChE..65.1710Z} provide a review of how these equations can be solved in their generic nonlinear form.
Computing nonlinear MHS requires accurate photospheric magnetic vector field measurements, which are not available in the quiet Sun regions and coronal holes due to instrumental limitations.

For the study of loops in the quiet Sun, only the longitudinal magnetic field component $B_l$ is measured accurately. Since the analyzed data are close to the Sun disk centre ( field of view of  400\arcsec\ $\times$ 400\arcsec, 1\arcsec$\sim$720~km), we neglect the small projection angle and assume that the vertical magnetic field component $B_z= B_l$.
We, therefore, solve the special class of LMHS equations derived by \citet{1991ApJ...370..427L} by assuming that the electric current density can be written in the form: 
\begin{equation}
\nabla \times {\bf B } = \alpha\, {\bf B } + a \exp(-\kappa z)\, \nabla B_z \times {\bf e_z}.
\label{def_j}
\end{equation}

This LMHS equation has three free parameters $\alpha, a,$ and $\kappa$.
The parameter $\alpha$ defines the strength of the field-aligned electric currents (like for a force-free field). The parameter $a$ is a pure number which measures the strength of the horizontal currents, and $1/\kappa$ prescribes the scale height of these currents. 
We compute the solution with the help of a Fast Fourier Transform (FFT). 
This technique was previously used to model magnetic structures where the plasma pressure and gravity deform the magnetic field such as within filaments \citep{Aulanier:1999}, in surges and arch filament systems \citep{Mandrini:2002}. For the required flux balance in the FFT method, we compute mirror magnetograms as introduced by \citet{1978SoPh...58..215S}. 
The original magnetogram, in the range $x=0 \dots L_x, y= 0 \dots L_y$, is mirrored to the regions with $x<0$ or/and $y<0$ to fill the region  $x=-Lx \dots L_x, y= -L_y \dots L_y$. The mirroring assumed that $B_z(-x,y,0)=-B_z(x,y,0)$,  $B_z(x,-y,0)=-B_z(x,y,0)$, and $B_z(-x,-y,0)=B_z(x,y,0)$ (where $x>0, y>0$). Then, the final magnetogram, twice larger in $x$ and $y$ directions, is flux-balanced by construction.

The FFT method solves the LMHS for given free parameters $\alpha$, $a$, and $\kappa$.
These three parameters are apriori unknown
and need to be computed from additional observations, here coronal images.
To do so we apply the recently developed algorithm
to compute the optimum LMHS parameters by comparing closed magnetic field lines with plasma loops
as seen in images. The main idea of this approach
is to compare closed magnetic field lines (which
are an output of the MHS code with arbitrary parameters)
with the emissivity seen in coronal images (see Paper~II for details). 

Our method is a generalization of a method developed in
\citet{Green:2002}, then in \citet{Carcedo_etal:2003}, where the optimum linear
force-free parameter $\alpha$ was computed by scanning the whole
parameter space and field lines were computed from footpoints limited to a manually selected area. 
Here, our approach rather selects the footpoint areas automatically
and chooses, as a standard option, the strongest positive and negative
magnetic elements from the magnetogram. 
All field lines connecting these footpoint areas are projected onto the EUV image.
Then, for each field line, we apply the following procedure:
First, we extract the EUV emissivity from AIA data along the magnetic field line on a band of $\pm$ 3 pixels perpendicular to it.  The lateral extension is limited to minimise the inclusion of  emission other than the EUV loop emission which will be selected during the minimisation process.   In the frame of the field line curvilinear abscissa and its orthogonal direction, the emissivity is located within an elongated rectangle with the selected field line located by definition along the central axis (as shown in Fig.~\ref{fig1}).  
At each location, $j$ ($=0$ to $m-1$) along the field line, the emissivity in the perpendicular direction is fitted with a Gaussian function. The shift of the Gaussian called $N_{\rm max}$, provides a measure of the distance between the field line and the fitted EUV loop.  How well the full field line agrees with the loop
in EUV is quantified with
  \begin{equation}
  C_i^2(\alpha,a,\kappa) =
  \sum\limits_{j=0}^{m} \frac{N_{{\rm max}, j}^2}{m-1}.
  \label{def_ci}
  \end{equation}

The quantity $C_i^2$ has to be minimized over all the computed field lines with respect
to the model parameters as
  \begin{equation}
  C_{\rm MHS}(\alpha,a,\kappa) = {\rm Min} (C_i^2 (\alpha,a,\kappa)).
  \label{def_cMHS}
  \end{equation}
This defines $\alpha$, $a$, and $\kappa$ for the studied region containing the loop, as well as the field line which best matches the EUV loop.
Finally, the method creates a so-called uncurled EUV image when convergence is achieved. Examples are shown in Fig.~\ref{fig1}. 

A further sophistication is that in images with several close-by EUV loops, we aim to select the brightest loop. This is done by defining
  \begin{equation}
  L_{\rm MHS}(\alpha,a,\kappa)=C_{\rm MHS} \cdot I_{\rm uncurled}^{-n},
  \label{defL}
  \end{equation}
where $I_{\rm uncurled}$ is the observed loop intensity, integrated along and across the uncurled loop and normalized to the average emissivity of the image.
In the present study, we choose $n=1$. This functional
$L_{\rm MHS}$ has to be minimized to find the
optimum model parameters $\alpha$, $a$, $\kappa$,
which is done with the help of a simplex-downhill method
as defined in \citet{Wiegelmann_Madjarska:2023}, section 6 `fully automatic method'. 
To save computing time, we evaluated in this first step only magnetic field lines which magnetically connect the strongest positive and
strongest negative magnetic elements in the FOV. We refer to it as the standard approach. 

A visual inspection of the resulting magnetic loops revealed that in several cases the loops visible in the AIA~193 images did not connect the strongest negative and positive flux concentrations in the 
HMI image data. We recomputed these cases with other magnetic
element pairs, selected by hand. We refer to it as a non-standard approach.  The photospheric magnetic field of the loops analysed here is far above the HMI errors of 10~G, and in the present case 4~G, as we used 8 binned  magnetograms, therefore the extrapolation accuracy is highly reliable.

\begin{figure}[!t]
    \centering
    \mbox{\includegraphics[width=3.7cm]{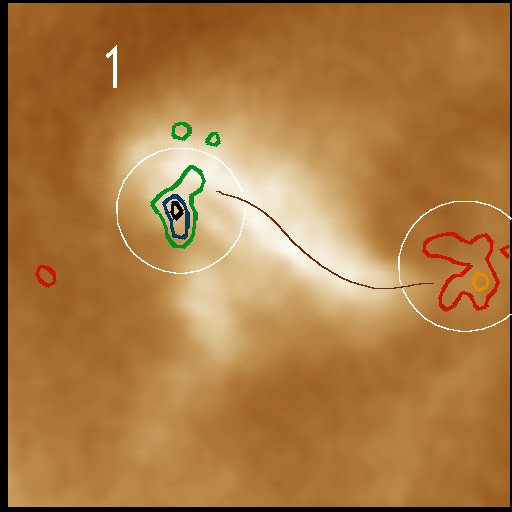}
    \includegraphics[width=4.1cm]{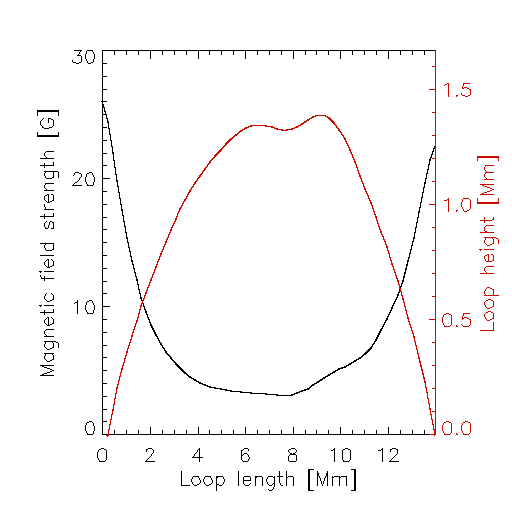}}
    \mbox{\includegraphics[width=3.7cm]{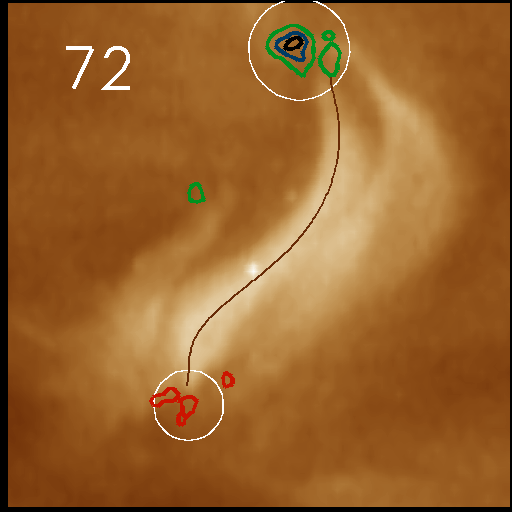}
    \includegraphics[width=4.1cm]{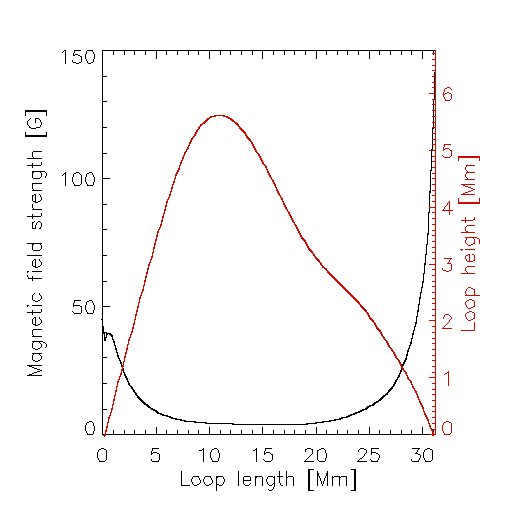}}
    \caption{Two examples of loops fitted with the fully (standard) automatic algorithm (cases 1 and 72). 
    {\it Left:} AIA~193 intensity with the best field line (black line) fitted a loop. The circles mark the potential footpoint areas as automatically selected by the algorithm. Only closed magnetic field lines 
    with footpoints in these circled areas are considered for the fitting procedure.
    The three black and green contours indicate negative polarities. The three orange and red contours outline positive polarities. All contours are relative to the absolute maximum field strength in the magnetograms, equispaced between 0 and max $|B_z|$. The intensity of each image is scaled to the maximum intensity of the image.
    {\it Right:} The black curve shows the magnetic field strength along the loop length as obtained from the LMHS model. The red curve  
    gives the loop height along its length. }
    \label{fig2}
\end{figure}

\begin{figure}[!t]
    \centering
    \mbox{\includegraphics[width=3.7cm]{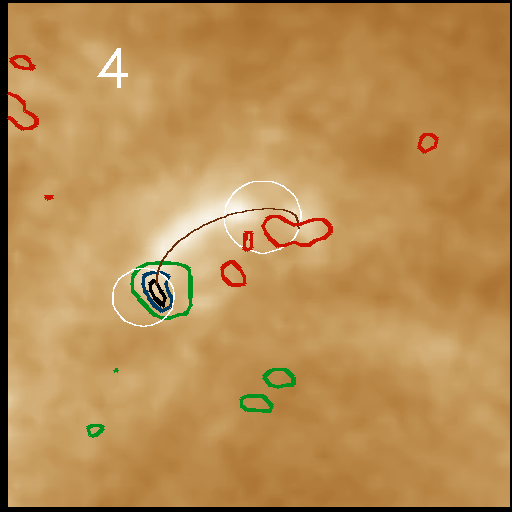}
    \includegraphics[width=4.1cm]{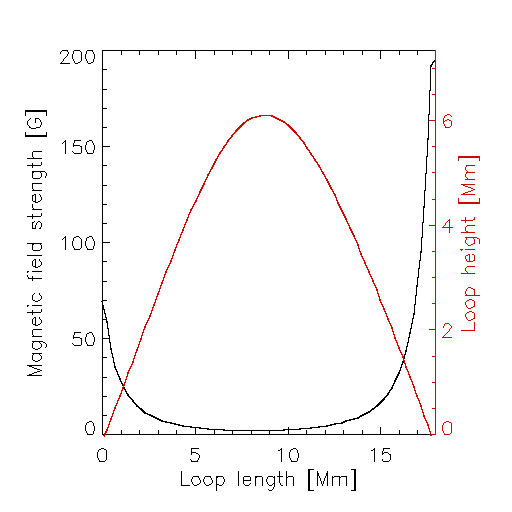}}
    \mbox{\includegraphics[width=3.7cm]{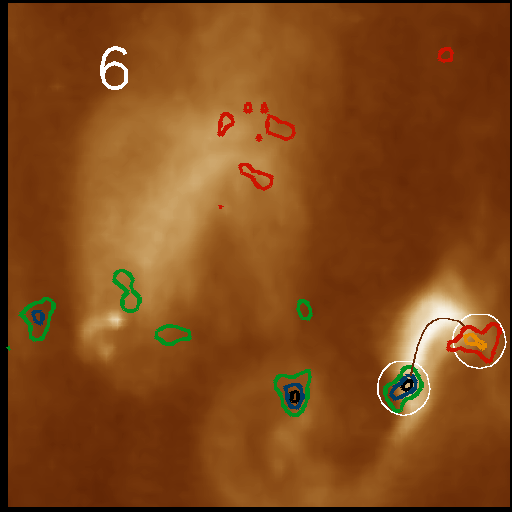}
    \includegraphics[width=4.1cm]{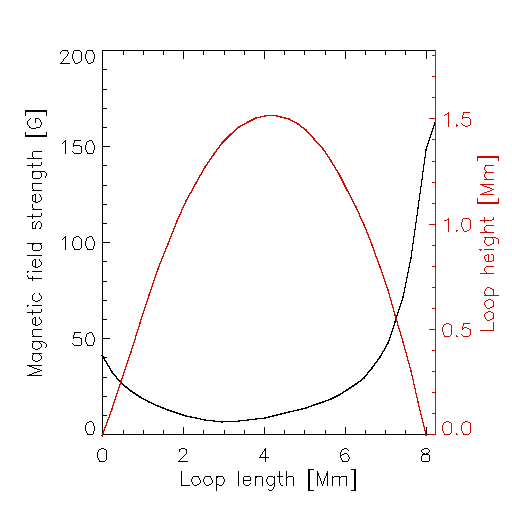}}
    \caption{Same as Fig.~\ref{fig2} of cases 4 and 6 for field lines obtained using a non-standard approach. See Section~\ref{Global} for details.
    }
    \label{fig3}
\end{figure}

\section{Results and Discussion}
\label{res}

\subsection{Global description}
\label{Global}
For this study, we randomly chose 189 frames containing small-scale loop systems (SSLSs), some bright, others faint, as selected in Paper~II.  The challenges of using all 189 frames come from the fact that not all loops can be fitted with the automatic algorithm, and finding the best parameters is time-consuming. Of those, in 126 frames we found a good agreement between a magnetic field line and a coronal loop seen in the Fe~{\sc xii}~193~\AA\ channel of AIA (hereafter AIA~193). In the first column of Table~\ref{table2}, we list each SSLS as bpXXX. where XXX is the numeration as in Paper~I. Eleven of these loops were extracted from faint SSLSs as noted in Paper~I. We should remark that some of the other loops are also fainter although composing generally bright SSLSs, i.e. a CBP.
When several loops are present in the same frame, the procedure selects the brightest one. 
Some of the 126 loops were selected in the same loop system to test the temporal variation of $\alpha$, $a$, and $\kappa$ as well as the topological and quantitative properties of different loops in the same SSLSs or the temporal changes of the same loop. Concerning the latter, a dedicated study is in progress using 45~sec cadence AIA and HMI data. These loops are often taken in consecutive or close-by-in-time images. 

For 97 out of 126 loops (77\%), the modelling is done with the standard approach, 
i.e. the loops connect the strongest positive and strongest negative magnetic elements. The remaining 29 loops (23\%) are obtained with a non-standard approach.
This is noted in column~6 (named App) of Table~\ref{table2}, where `St' refers to parameters obtained with a standard approach and `NSt' to those obtained with a non-standard approach. 
In Fig.~\ref{fig1} we present the uncurled loops of all successfully extrapolated loop examples (126) sorted by their lengths. It can be noted that longer loops tend to be fainter. 

\begin{figure*}[!ht]
    \centering
    \includegraphics[scale=0.4]{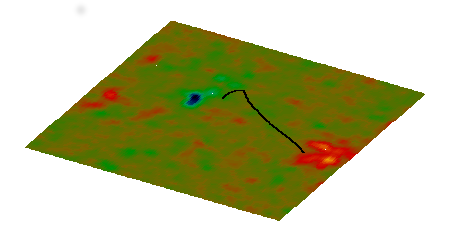}
    \includegraphics[scale=0.4]{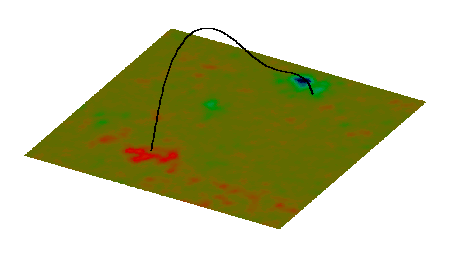}\\
    \includegraphics[scale=0.4]{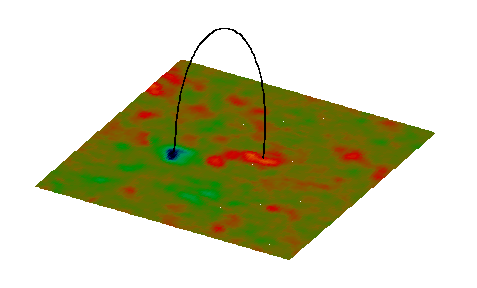}
    \includegraphics[scale=0.4]{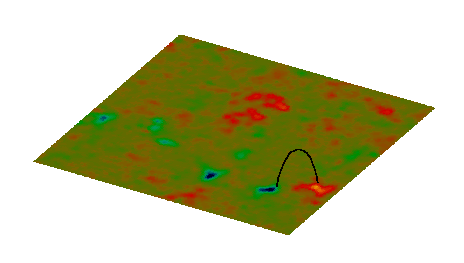}
    \caption{Four examples of 3D views for the best-computed field line (black line) matching the best observed AIA~193 loop. The observed magnetogram is shown at the bottom of the 3D plots with green/red colours for negative/positive $B_z$ values, respectively.
    {\it Top row:} HMI magnetogram of cases 1 and 72 shown in Fig.~\ref{fig2}.
    {\it Bottom row:} Same for cases 4 and 6 
    shown in Fig.~\ref{fig3}.}
    \label{fig_3D}
\end{figure*}

In Fig.~\ref{fig2} two examples of loops obtained with the fully automatic algorithm (standard) are shown, while Fig.~\ref{fig3} presents two examples of the non-standard approach. A 3D view of all four loops is presented in Fig.~\ref{fig_3D}. The rest of the images (as in the left panels in Figs.~\ref{fig2} and \ref{fig3}) are shown in Figs.~\ref{app-fig1}--\ref{app-fig3}. Finally, the detailed results obtained for the 126 loops are given in Table~\ref{table2}, including the parameters $\alpha$, $a$ and $\kappa$ in columns, 3, 4 and 5; the averaged intensity (I, column~7), the loop lengths (L, column~8), height (H, column 9), the magnetic flux in the positive and negative footpoints of the loop (B$_{pos}$ and B$_{neg}$, columns 10 and 11) and the average magnetic flux along the loop (B$_{\rm av}$, in column 12). In Paper~I we provide the animations (intensity and photospheric magnetic field evolution) of all SSLSs studied there\footnote{\url{https://zenodo.org/record/8163995}}.

\begin{table}
\caption{Statistics of the morphological and magnetic properties of small-scale loops. 
The following acronyms are used: L -- length, H -- height, Av. -- average, and Med. -- median, $B_{\rm av}$ -- average magnetic field along the loop, $B_{top}$ --  magnetic field at loop top, $B_{\rm weak}$ -- weak magnetic field footpoint, $B_{\rm strong}$ -- strong magnetic field footpoint. }
\begin{center}
\begin{tabular}{ccccc}
\hline
\hline
      & All loops & H $\leq$ 1.5 Mm & H $>$ 1.5 Mm \\
\hline
Number &  126 &   38 &   88\\
Med. |$\alpha$|&  2.2 &  2.2 &  2.2\\
Av. |$\alpha$| &  2.4 &  2.4 &  2.4\\
Med. $a$&  0.7 &  0.9 &  0.7\\
Av. $a$ &  0.6 &  0.7 &  0.6\\
\hline
Max loop H (Mm) & 13 &  1.4 & 13\\
Min loop H (Mm) &  0.2 &  0.2 &  1.5\\
Av. loop H (Mm) &  4 &  1 &  5\\
Max loop L (Mm) & 60 & 14 & 60\\
Min loop L (Mm) &  1.4 &  1.4 &  6\\
Av. loop L (Mm)& 17 &  8 & 20\\

Med. L~/~H &  5 &  9 &  4\\
Av. L~/~H &  7 & 12 &  5\\
\hline
Max |$B_{\rm av}$| (G)  & 81 & 66 & 81\\
Min |$B_{\rm av}$| (G) &  5 &  8 &  5\\
Av. |$B_{\rm av}$| (G) & 23 & 25 & 23\\
\hline
Max |$B_{top}$| (G) & 60 & 60 & 33\\
Min |$B_{top}$|  (G) &  1 &  4 &  1\\
Av. |$B_{top}$|  (G) & 10 & 17 &  8\\
\hline
Max |$B_{\rm strong}$| (G) &504 &147 &504\\
Min |$B_{\rm strong}$| (G) &  4 &  4 & 12\\
Av. |$B_{\rm strong}$| (G) &112 & 50 &138\\
\hline
Max |$B_{\rm weak}$| (G) &244 & 79 &244\\
Min  |$B_{\rm weak}$| (G) &  0.3 &  1 &  0.3\\
Av.  |$B_{\rm weak}$| (G) & 47 & 22 & 58\\
\hline
\hline
\end{tabular}
\end{center}
\label{table1}
\end{table}

\begin{figure*}[!ht]
\centering
\includegraphics[height=7cm]{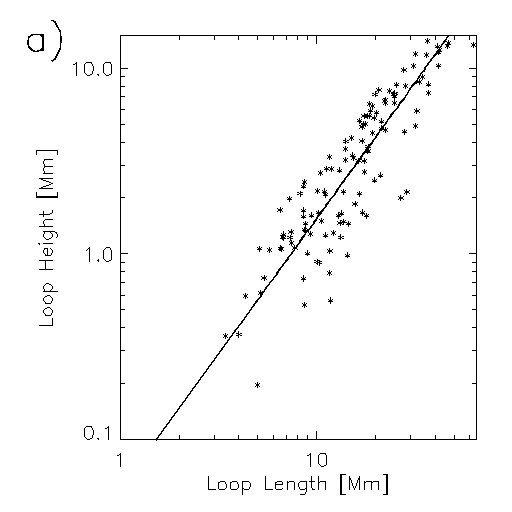} 
\includegraphics[height=7cm]{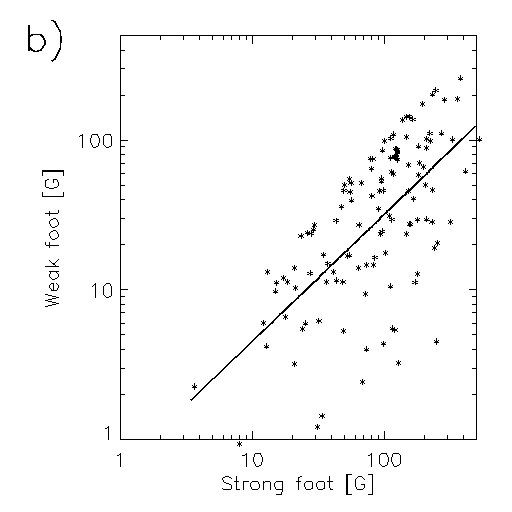}\\
\includegraphics[height=7cm]{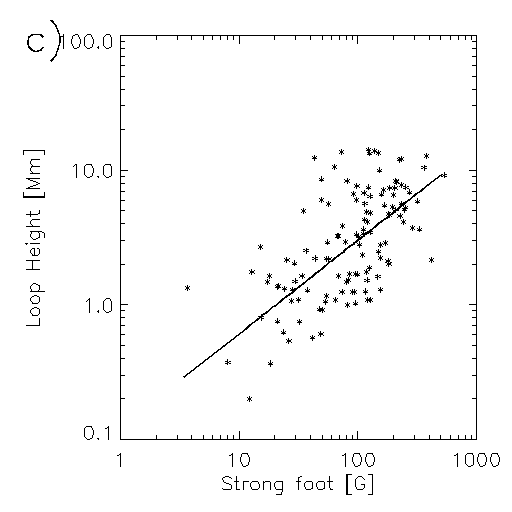}  
\includegraphics[height=7cm]{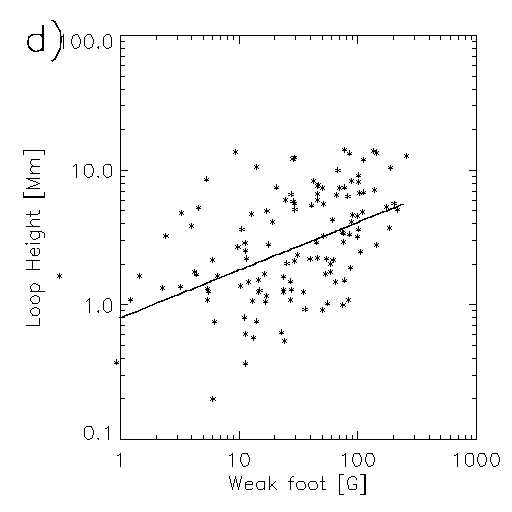}\\
\includegraphics[height=7cm]{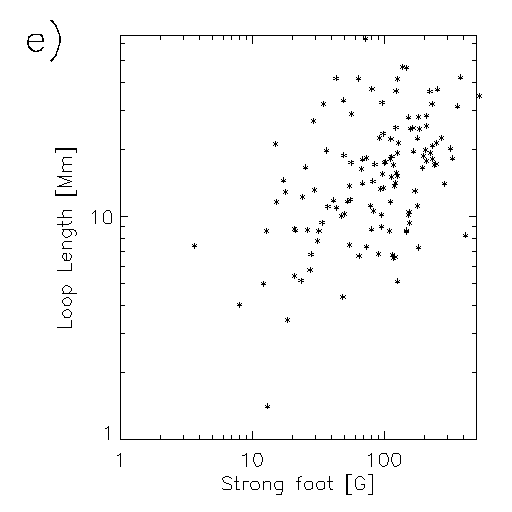}  
\includegraphics[height=7cm]{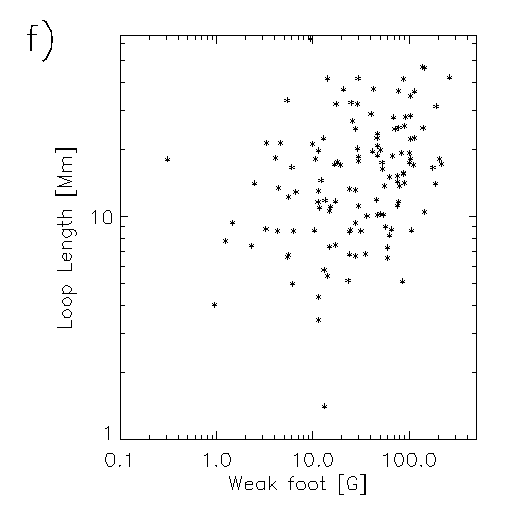}
\caption{Scatter plots of some statistical properties for 126 loops.  
    a) Loop length vs. loop height (Pearson correlation $c_{\rm P}=$ 0.90), 
    b) Magnetic field strength in the strong vs. weak footpoint of magnetic loop ($c_{\rm P}=$ 0.60),  
    c) Stronger magnetic field footpoint vs loop height ($c_{\rm P}=$ 0.46),  
    d) Weaker magnetic field footpoint vs. loop height ($c_{\rm P}=$ 0.45), 
    e) Stronger magnetic field footpoint vs. loop length ($c_{\rm P}=$ 0.34), and 
    f) Weaker magnetic field footpoint vs. loop length ($c_{\rm P}=$ 0.31).
    }
    \label{fig5}
\end{figure*}

\begin{figure*}[!]
    \centering
\includegraphics[height=7cm]{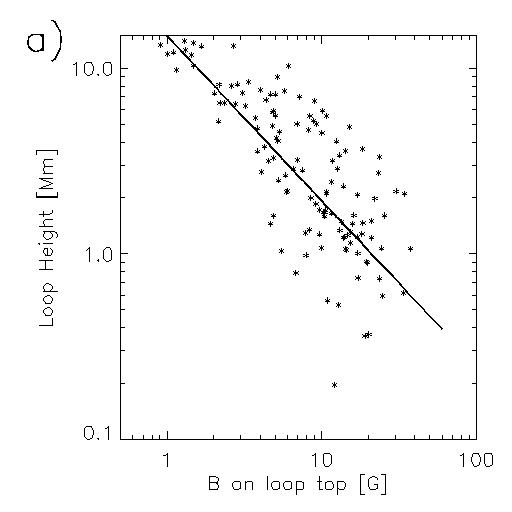} 
\includegraphics[height=7cm]{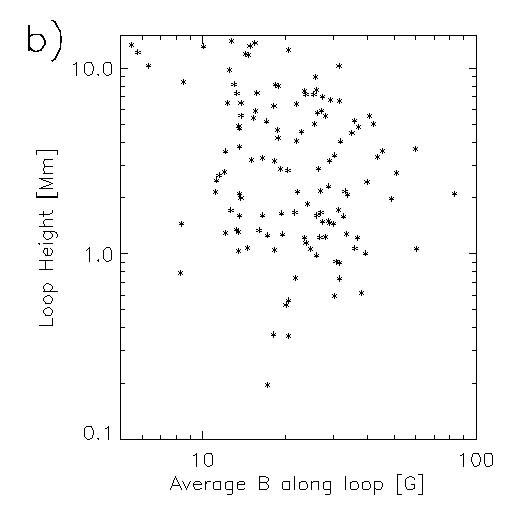}
\includegraphics[height=7cm]{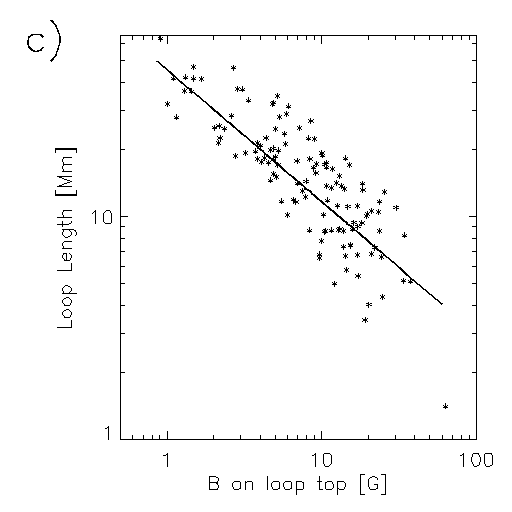}
\includegraphics[height=7cm]{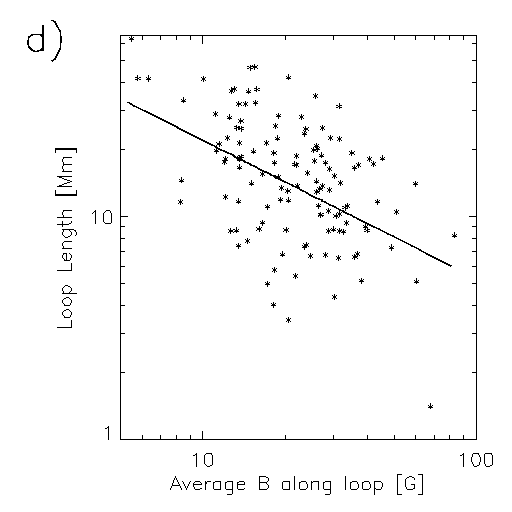}
\includegraphics[height=7cm]{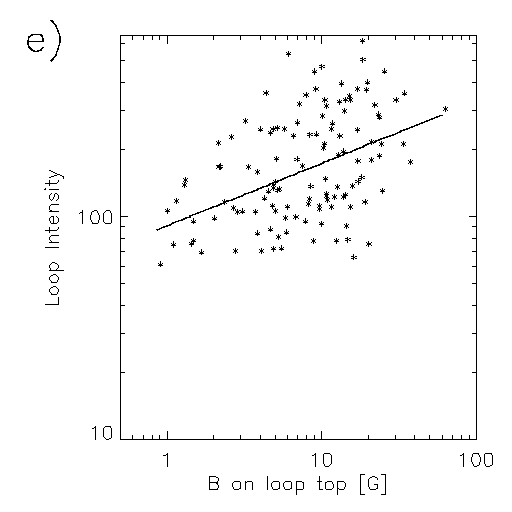} 
\includegraphics[height=7cm]{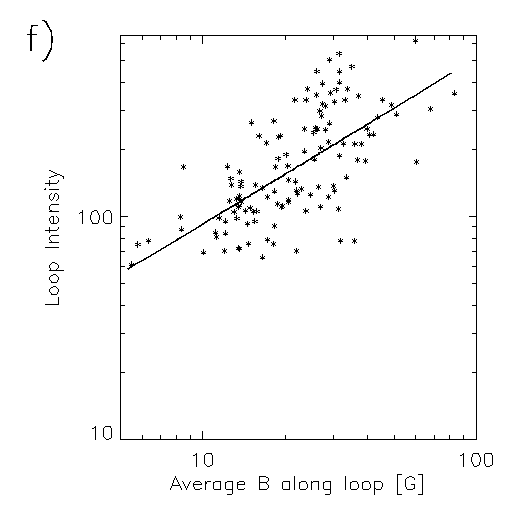} 
\caption{ Scatter plots of some statistical properties for 126 loops. a) Magnetic field strength at the loop top vs. loop height (Pearson correlation coefficient $c_{\rm P}$ of $-0.59$),  
    b) Average magnetic field strength along the loop vs. loop height ($c_{\rm P}=-0.30$)
    c) Magnetic field strength  on loop top vs. loop length ($c_{\rm P} =-0.64$), 
    d) Average magnetic field strength along the loop vs. loop length ($c_{\rm P}=-0.44$), 
    e) Magnetic field strength on loop top vs. loop average intensity ($c_{\rm P}=0.35$), 
    f) Average magnetic field strength along the loop top vs. loop average intensity ($c_{\rm P}=0.58$). }
    \label{fig6}
\end{figure*}

\subsection{LMHS parameters}
\label{lmhspar}
In the second column of Table~\ref{table2}, we provide the functional $L_{\rm MHS}$ for all 126 loops. The smaller the value of $L_{\rm MHS}$ in Eq.~(\ref{defL}), the better the magnetic field lines and AIA~193 loops agree. For 94 out of the 126 loops, we get $L_{\rm MHS} <1$ (see Paper~II for more details about the functional $L_{\rm MHS}$).
Visual inspection confirms the excellent agreement of the magnetic-field lines obtained from the LMHS modelling and the loops seen in the AIA~193~\AA\ channel ($\log T \sim$ 1~MK). The optimum solution is when $L_{\rm MHS}$ is equal to zero (for more details see Paper~II). Higher values of $L_{\rm MHS}$ correspond to worse agreement between a field line and a loop seen in AIA~193.
The reason for that is that at some points along the field line,
the maximum of the EUV emissivity is shifted away from  
the field lines, leading to higher values in Eq.~(\ref{def_ci}), e.g. cases 24 and 39 (see Fig.~\ref{app-fig1}).
Higher values in $L_{\rm MHS} $ can also occur for very faint plasma loops  when $I_{\rm uncurled}$ in 
the functional of Eq.~(\ref{defL}) becomes very small, e.g. cases~85  and 87 (see Fig.~\ref{app-fig3}).
 
As mentioned above, in columns 3--5 of Table~\ref{table2} annotated with $\alpha$, $a$, 
and $\kappa$, we show the corresponding optimized parameters of
the LMHS model as defined in Eq.~(\ref{def_j}). 
As mentioned in Section~\ref{method}, the parameter $\alpha$ controls the field-aligned electric current density. The higher the absolute value of $\alpha$ the stronger the electric current is. 
For positive values of $\alpha$, the currents are parallel to
the magnetic field and for negative values of $\alpha$, 
the currents are antiparallel to the magnetic field.
The parameter $a$ controls the horizontal currents,
which are partly perpendicular to the magnetic field lines
and cause a finite Lorentz force. Large values of $a$ 
mean that the magnetic structure is far away from a force-free equilibrium. 
Finally, $\kappa$ defines how fast the non-magnetic forces
decrease with height. 
If the parameter $a=0$, we have no Lorentz force
and the LMHS model reduces to a linear force-free model,
which is the case for 14 cases. 

A subclass of linear force-free equilibria are potential fields where also
$\alpha=0$. This is not the case in any of the investigated examples,
except for case 2 where $\alpha=0.01$ so that the magnetic field is nearly potential.
The $\alpha$ values reported in Table~\ref{table2} are normalized by the magnetogram size. 
Out of all 126 loops, only 32 have a value of $ |\alpha| < 1.0$ and 17 of them have $ |\alpha| < 0.5$.  These loops have rather moderate field-aligned electric currents, 
i.e. the field is close to potential for these cases. 
All other force-free loops have absolute values of $|\alpha|$ up to 6  and contain 
rather strong field-aligned electric currents. 

 The parameter $a$ is a dimensionless number in Eq.~(\ref{def_j}) which is limited in the model to $a \leq 1.2$. Only 
10 loops have $a=0$. Moreover, only one loop (case 20) has both $a=0$ and a small value of $\alpha=-0.14$, so it is very close to a potential magnetic field. 

The ratio $1/\kappa$ is the height at which the non-magnetic forces drop to $1/e$ (about 37\%).
This can be assumed as the height of the forced layer. In Table~\ref{table2} $1/\kappa$ is given in Mm$^{-1}$.  
For instance, for a pixel size of 432~km as in the present work, the forced layer has a height of 2.2~Mm for $\kappa=0.2$ and a height of 0.72~Mm for $\kappa=0.6$. If $a$ is equal to zero, we have a force-free field and therefore no $\kappa$.  The values of $1/\kappa$ are typically small and the median is only 0.7 Mm, so about half of the vertical extension of the classical chromosphere. The median of loop height is also small, 2.7~Mm, while still above a factor 3 larger than the median of $1/\kappa$. This means that, apart from their footpoints, most loops have nearly a force-free field in the corona above the forced layer.

Cases  25 and  26 correspond to the same loop recorded 6~min apart. One could note that the parameter $\alpha$ has changed from $-0.16$ to $0.16$ which can be considered negligible. The same is valid for all other parameters (see Table~\ref{table2}). The same can be seen for cases 43 and 44  where the images are taken 12~min apart. Cases 110--115 sample different loops in the same CBP, BP046, with a time difference between the frames from 12~min to 14~h (the time difference between two frames is 6~min as explained in Paper~I).  The loops are found to have a different $\alpha$ parameter, ranging from $-0.8$ to $5.5$ indicating the energetic complexity of small-scale systems as in ARs \citep[e.g., ][]{2014LRSP...11....4R}{}{}.  

We learned from the investigation in Paper~II, that $\alpha$ influences the loop-fitting more than $a$ and $\kappa$. This is linked to the vertical small extension of the forced layer compared to the loop height (so a large fraction of loops are typically force-free). 
As tested in Paper~II, it is possible to do a constraint optimization (fix $a$ and $\kappa$) and vary only $\alpha$. From the few examples in Paper~II, the fitting (values of $L_{mhs}$) is only slightly worse than fitting all three parameters. This could be further explored in the future if we have for instance vector magnetograms from which we could deduce $a$ from measurements at photospheric and chromospheric heights (e.g. with photospheric and chromospheric vector magnetograms).
Furthermore, we could compute the Lorentz force and
the related parameter $a$ in both photospheric and chromospheric magnetograms to find out how fast $a$ drops with height from the photosphere.

\subsection{Loop physical parameters}
\label{looppar}

Table~\ref{table1} summarizes some of the physical parameters obtained from the LMHS model. The loops' height (H) range is from 0.2~Mm (we exclude here case 28 which has a height of 0.01~Mm) to 13~Mm. We find that 38 loops have heights below 1.5~Mm (the typical height of the chromosphere) and 88 loops above 1.5~Mm. The average height taking into account all 126 loops is 4~Mm. Loop lengths are in the range from 1.4~Mm to 60~Mm with a median of 15 Mm. The average loop length for loops below typical chromospheric heights of 1.5~Mm is 8~Mm, while above this height is 20~Mm. For comparison, loop heights measured in active regions are typically in the range of 10--93~Mm and have loop lengths in the range from 60 to 350~Mm \citep[e.g.][]{2008ApJ...679..827A, 2017ApJ...842...38X}. Thus, the loops studied here are significantly less tall and shorter than AR loops. 

A natural question to address is whether the field lines with low heights, i.e. as low as 0.2~Mm,  indeed confine plasmas heated to a million degrees. As a matter of fact, the observations show very short bright elongated EUV features connecting magnetic flux concentrations of opposite polarities. Those short bright loops are typically observed towards the end of the lifetime of SSLSs (see the animation of Paper~I). To further explore the truthfulness  of this observation, that indeed small-low lying loops confining million-degree plasma are present, one can employ spectroscopic and imaging co-observations as done in section~4 of the study by \citet{2008A&A...482..273M}. Such data permit investigation of whether the emission in the imaging channels comes from plasma at transition region or even chromospheric temperatures rather than coronal. It is well known that coronal EUV imaging channels could be contaminated with such emission, including SDO/AIA~193 \citep[for details see][and references therein]{2018A&A...619A..55M, 2022ApJ...929..103T}.   A forthcoming study will employ data from the IRIS, SDO/AIA and Solar Orbiter Extreme-Ultraviolet Imager (EUI) to explore this in full detail.  

The loops studied here have a median (average) length/height ratio of 9 (12) for the loops with heights below 1.5~Mm and 4 (5) for loops higher than 1.5~Mm. The median (average) length/height ratio for all 126 loops is 5 (7).  This indicates that lower loops are significantly flatter than a half-circle loop (length/height $=\pi$) while higher loops are more rounded.
The flatness of lower loops is due to the significant role of non-magnetic forces including gravity and plasma pressure at heights below 1.5~Mm (classical chromospheric top height). 

The average magnetic field strength $B$ along the loops when all loops are considered ranges from 5~G to 81~G. Loops with heights below 1.5~Mm  reach maximum values of $B$ of 66~G while $B$ can be as high as 81~G  for loops higher than 1.5~Mm. However, if we consider the average values, they indicate that loops lower than 1.5~Mm have similar average magnetic-field strength along the loops than the higher loops (25~G compared to 23~G, respectively).  

The average magnetic field $B$ at loop tops for lower loops ($<$1.5~Mm) is 17~G compared to 8~G for the higher loops (heights above 1.5~Mm), i.e. a factor 2 lower.  These are mean tendencies within a broad range of values from 1~G to 60~G for all the loops. For lower loops (heights below 1.5~Mm), the maximum of $B$ is twice higher (60~G) than for loops above 1.5~Mm (33~G), so the same result as above with the average magnetic field $B$ at loop tops. 

Next, we compare the footpoint magnetic field measure from the starting(end)-point to a field line.
The magnetic field of the loop footpoints is from $\sim$500~G  (max $B_{\rm strong}$) to as weak as $\sim$0.3~G (min $B_{\rm weak}$).  Lower loops are rooted in a weaker magnetic field with an average $B_{\rm weak}$/$B_{\rm strong}$ of 22/50~G, while for higher loops the values are  58/138~G, i.e. twice higher. The difference in the magnetic field strength in the two footpoints may cause siphon flow \citep{2012A&A...537A.130B}. However, to establish this, spectroscopic data are needed to obtain the Doppler velocities. This will be the subject of another future statistical work where  Interface Region Imaging Spectrograph (IRIS)  data will be employed.

The inclination of the magnetic field (defined as the angle between $\vec{B}$ and the vertical z-axis)  was also explored. We find some correlation between the inclination at the weak and strong $B$ footpoints with a Pearson correlation coefficient $c_{\rm P}=0.52$ (Fig.~\ref{app-fig4}, left panel). The inclination in both footpoints anti-correlates 
with loop heights with the same Pearson coefficient $c_{\rm P}$ = $-0.47$  (Fig.~\ref{app-fig4}, middle and right panels). Indeed, it is expected that higher loops have more vertical footpoints.  Very weak anti-correlation is found for the strong ($c_{\rm P}=-0.28$) and weak ($c_{\rm P}=-0.25$) magnetic field footpoints and loop length.

\subsection{Relationship of loop physical parameters}
\label{relation}

Figures~\ref{fig5} and \ref{fig6} provide the statistical correlation for the various loop parameters. 
The highest correlation is between the lengths and heights of the loops with a Pearson correlation coefficient $c_{\rm P}=0.9$, Fig.~\ref{fig5}a. We obtained a power law index $\gamma$ =  1.46 for the linear fit $log(H) = \gamma \, \log (L) - 1.3$. 
One can notice that loops with lower heights deviate asymmetrically from the linear fit, possibly related to the influence of gravity and pressure at lower heights. 

We find a relatively high correlation ($c_{\rm P}=0.6$) between the magnetic field strength in the two footpoints (Fig.~\ref{fig5}b). The relationship is a power law of $\log (B_{\rm weak}) = 0.85\, \log (B_{\rm strong}) - 0.2$. 
There is some, although not a strong correlation between the footpoints with stronger and weaker magnetic fields  and loop heights ($c_{\rm P}=$ 0.46 and 0.45, respectively, (Fig.~\ref{fig5}c, d). For the power law relationships, we obtained $ \log ({H}) = 0.7\, \log (B_{\rm strong}) - 0.9$ and $ \log ({H}) = 0.36\, \log (B_{\rm weak}) - 0.1$, respectively. 
The correlation is weaker for the magnetic field at footpoints versus loop lengths, $c_{\rm P}$ of 0.34 and 0.31, respectively (Fig.~\ref{fig5}e, f). 
The two relationships could be represented in average with the power laws of $\log ({L})  = 0.32\, \log (B_{\rm strong})  + 0.5$ and $ \log ({L}) = 0.16\, \log (B_{\rm weak})  + 0.9$, respectively.
 
Next, we describe the correlations between the magnetic field strength at the loop top and its mean along the loop. First, we notice that the loop with the smallest length and height is isolated from the other loops (Fig.~\ref{fig6}a, c, d), therefore, we exclude it from our statistics. Next, there is a clear anti-correlation between the field strength on the loop tops ($B_{top}$) and heights ($c_{\rm P}=-0.59$, Fig.~\ref{fig6}a). The relationship is a power law of $ \log (H) = -0.89\,\log (B_{top}) + 1.2$. As in Fig.~\ref{fig5}a, the smaller loop heights (below $\sim$2~Mm) deviate asymmetrically from the linear fit in Fig.~\ref{fig6}a.  A slightly stronger correlation is present with loop lengths ($c_{\rm P}=-0.64$, Fig.~\ref{fig6}c) and the power law relation is $ \log (L)= -0.59\, \log (B_{top})  + 1.7$. 
In contrast, we find a weak correlation between the average magnetic field strength along the loop $B_{\rm av}$ and the loop heights and lengths with  $c_{\rm P}$ = $-0.3$ and $-0.44$, Fig.~\ref{fig6}b and d, respectively.  The two relationships are a power law of $\log (B_{\rm av}) = -0.63\, \log ({H}) + 1.2$ and $\log (B_{\rm av)} = -0.62\, \log ({L}) + 2.0$, respectively.

This contrasts with the results of \citet{Mandrini:2000} obtained for coronal loops within active regions (ARs) where $c_{\rm P}=-0.88$ between the loop $B_{\rm av}$ and lengths.  However this result is obtained in different conditions: within ARs, with $100 \leq B_{\rm foot} \leq 500$~G and for longer loops ($50 \leq L \leq 300$ Mm). They found that for $L$ below 50 Mm, $B_{\rm av}$ is almost independent of loop length in ARs.  Here, a similar result is obtained in the quiet Sun.
 
Next, we investigate the correlation with the loop intensity, which is defined as an average along the loop length with 1 pixel on both sides of the loop centre. The lengths and heights of the loops are uncorrelated with the loop intensity  ($c_{\rm P}$ of $-0.19$ and $-0.12$, respectively). 
The intensity  of the loop is correlated ($c_{\rm P} = 0.58$, $\log (I) = 0.74\log (B_{av}) + 1.2$) with the average magnetic field strength along the loop (Fig.~\ref{fig6}f). This correlation is stronger than the intensity correlation with the field strength on the loop tops ($c_{\rm P} = 0.35$,  Fig.~\ref{fig6}e).  The relationship is a power law of $ \log ({I}) = 0.28\, \log (B_{\rm top})  + 2.0$. These correlations indicate that the energy release in the loop is more linked to the average $B$ along the loop than the field strength on the loop top. In other words, the energy is likely released all along the loops, not only at the loop tops.  This is well plausible, but it remains that the short length of the loops allows an efficient redistribution of the input energy along the loop.  More precisely, the thermal conduction could be so efficient in such short loops (it scales as $L^{-2}$) that it could transfer the energy input all along the loop, independently of its original input location. In comparison to X-ray loops in ARs, the results show a dispersion that is broadly similar to the present result. In figures~1 and 2 of \citet{Klimchuk_etal:1995},  the temperature and plasma pressure are anti-correlated with loop lengths but with a substantial dispersion.  \citet{Mandrini:1996} investigated the relationship between loop length and $B$ but only statistically (not for individual loops). $B$ was found to anti-correlate with loop lengths with also significant dispersion (Figure~2 in their paper). To conclude, we can speculate that the loop intensity, which is a function of plasma density and temperature, is correlated with $B$ but with a substantial dispersion.

This brings us to the discussion and future applications of the present study for the validation of present and future models of small-scale loops
that confine plasma heated to a million degrees. The most recent are the 3D radiative MHD models by \citet{2023ApJ...958L..38N} and \cite{2022ApJ...937...91C}. While the former is based on the Bifrost code \citep{2011A&A...531A.154G}, the latter employs the MURaM code \citep{2005A&A...429..335V, 2017ApJ...834...10R}.   Previously, all CBP models were built on ad-hoc driving mechanisms, and magnetic reconnection was employed  as the main heating mechanism.  One pioneering model for its time is the so-called Converging Flux Model (CFM) \citep[e.g.][]{1994ApJ...427..459P, 2006MNRAS.369...43V} where the magnetic
energy is converted through magnetic reconnection in the corona to thermal
energy. More recently, \cite{2018ApJ...864..165W} developed a model exploring the formation of CBPs
in coronal holes and associated collimated flows, that is jets. 

The model by \citet{2023ApJ...958L..38N} has been specifically developed to understand the plasma heating of  small-scale loops  that are the subject of investigation in the present study. In this model, the energy injection is produced through surface convection.  The model explains the sustainability of the CBP heating for several hours. It also provides unprecedented observational diagnostics that are compared with simultaneous space and ground observations. The model is based on fan-spine configuration and demonstrates how stochastic motions associated with photospheric convection can significantly stress the CBP magnetic field topology, leading to important Joule and viscous heating around the CBP inner spine at a few megameters above the solar surface (transition region and low corona).  Magnetic reconnection in the null point in the corona has some but lesser (than the Joule and viscous) heating contribution. The results from the present study will be directly compared with this model to verify its validity and will be reported in a forthcoming study.

\section{Summary and conclusions}

We present here the first-of-its-kind statistical study on the coronal magnetic field properties of small-scale loops observed in emission with formation temperatures at $\sim$1~MK. The brighter of these loops compose the so-called coronal bright points. The study employs a recently developed automatic algorithm to compute 
LMHS equilibria that match computed field lines with enhanced emission features in the SDO/AIA~193 channel. 

We report the statistical properties of 126 loops for which we find a good agreement between a computed magnetic field line and an observed coronal loop seen in the Fe~{\sc xii}~193~\AA\ channel of AIA.  The magnetic field
is found to be non-potential with rare cases of field lines that are close to potential. Loops preserve the same $\alpha$ for a short time period which is at least 6~min.  A forthcoming dedicated study will explore the lifetime and temporal variation of the coronal magnetic properties of small-scale loops using a 45~s data cadence. 

We summarize here our main results. The average loop height is 4~Mm, while the average loop length is 17~Mm. Loops at lower heights, below 1.5~Mm (considered classic chromospheric height), are flatter possibly caused by the role of gravity and plasma pressure that are in play at these heights. The average magnetic field strength $B$ along the loops ranges from 5~G to 81~G. The magnetic field at loop tops varies from 1~G to 60~G.  The highest Pearson correlation coefficient ($c_{\rm P}$) is found between the lengths and heights of the loops ($c_{\rm P}$ = 0.9). 

The loop intensity correlates stronger with the average magnetic field along the loops than with the magnetic field at loop tops suggesting that the main energy deposition that causes the plasma to heat to up to a million degrees does not happen on loop tops but rather occurs at lower heights along 
the loops as also predicted by the recent model of \citet{2023ApJ...958L..38N}. Meanwhile, simulations with the MURaM code by \cite{2022ApJ...937...91C} have suggested magnetic braiding as a possible mechanism for the heating of small-scale loops without presenting a detailed investigation which is hopefully forthcoming.

The present study will be extended to data from the Solar Orbiter EUI high-resolution imager. Given its higher spatial resolution, it is likely that one AIA 193 loop will be composed of several strands. Therefore, the properties of the loop may depend on the number of strands and their degree of interlacement. At the photospheric level, it will be worthwhile to determine if the magnetic field ($B$) is monolithic or consists of several flux concentrations. 

\label{sum_concl}

\begin{acknowledgements}
We want to thank very much the referee for their helpful comments and suggestions.
MM and TW acknowledge DFG grants WI 3211/8-1 and WI 3211/8-2, project number 452856778. 
TW acknowledges DLR grant 50OC2301.
The HMI and AIA data are provided courtesy  of NASA/SDO  science teams. The HMI and AIA data have been retrieved using Stanford University's Joint Science Operations Centre/Science Data Processing Facility.  MM and KG thank the ISSI (Bern) for the support of the team ``Observation-Driven Modelling of Solar Phenomena''. 
\end{acknowledgements}

\bibliographystyle{aa}
\bibliography{ref}

\begin{appendix}

\section{Properties of small-scale loops}
\label{Properties_loops}

Table~\ref{table2} lists the obtained results for the 126 loops.
Column 1: the CBP number (same numeration as in Paper~I) and the selected frame number. 
Column 2: the minimal value of $L_{\rm MHS}$ (Eq.~(\ref{defL})).
Columns 3--5: optimized parameters $\alpha$, $a$, and 1/$\kappa$ values of the LMHS model defined in Eq.~(\ref{def_j}).  $\alpha$ is normalized by the magnetogram size. 
Column 6: `St' refers to parameters obtained with a standard approach and `NSt' to the ones obtained with a non-standard approach.
Column 7: Loop mean intensity, I.
Columns 8--9: loop length, L, and height, H.
Columns 10--12: Magnetic field strength in the positive and negative footpoints, and average field strength along the loop.

\onecolumn
\onecolumn
\begin{center}
\begin{longtable}{cccccccccccc}
\caption{Coronal magnetic properties of small-scale loops in the quiet Sun.} \\
 \hline 
  \multicolumn{1}{c}{Case} &\multicolumn{1}{c}{L$_{MHS}$} &\multicolumn{1}{c}{$\alpha$}
 &\multicolumn{1}{c}{$a$} &\multicolumn{1}{c}{$1/\kappa$} &\multicolumn{1}{c}{App} 
 &\multicolumn{1}{c}{I} &\multicolumn{1}{c}{L} &\multicolumn{1}{c}{H} 
 &\multicolumn{1}{c}{B$_{pos}$} &\multicolumn{1}{c}{B$_{neg}$} &\multicolumn{1}{c}{B$_{av}$} \\
 
  \multicolumn{1}{c}{} &\multicolumn{1}{c}{} &\multicolumn{1}{c}{}
 & \multicolumn{1}{c}{} &\multicolumn{1}{c}{Mm$^{-1}$} &\multicolumn{1}{c}{} 
 &\multicolumn{1}{c}{DN/s} &\multicolumn{1}{c}{Mm} &\multicolumn{1}{c}{Mm} 
 &\multicolumn{1}{c}{G} &\multicolumn{1}{c}{G}  &\multicolumn{1}{c}{G}   \\
\endfirsthead

 \multicolumn{12}{c} {{ \tablename\ \thetable{} -- Continued  from previous page}} \\
 \hline
\multicolumn{1}{c}{Case}&\multicolumn{1}{c}{L$_{MHS}$}& \multicolumn{1}{c}{$\alpha$}&\multicolumn{1}{c}{$a$}&\multicolumn{1}{c}{$1/\kappa$}&\multicolumn{1}{c}{App}&\multicolumn{1}{c}{I}&
\multicolumn{1}{c}{L}&\multicolumn{1}{c}{H}&\multicolumn{1}{c}{B$_{pos}$}&\multicolumn{1}{c}{B$_{neg}$} &\multicolumn{1}{c}{B$_{av}$} \\
  
  \multicolumn{1}{c}{}&\multicolumn{1}{c}{}& \multicolumn{1}{c}{}&\multicolumn{1}{c}{}&\multicolumn{1}{c}{Mm$^{-1}$}&\multicolumn{1}{c}{}&\multicolumn{1}{c}{DN/s}&\multicolumn{1}{c}{Mm}&\multicolumn{1}{c}{ Mm}&\multicolumn{1}{c}{G}&\multicolumn{1}{c}{G}&\multicolumn{1}{c}{G} 
 \\ \hline
\endhead
\hline
\multicolumn{11}{c}{Continued on next page}
\endfoot
 \endlastfoot
 \hline
1 bp001 fr111 &   0.5& $-0.6$ &  0.4 &  0.6&  St &    84 &    14.0 &     1.4 &    16 &   $-11$ &     8\\
2 bp002 fr095 &   0.01&  0.01 &  1.00 &  0.6&  St &   111 &     3.3 &     0.3 &    18 &   $-11$ &    20\\
3 bp002 fr134 &   0.01& $-0.4$ &  0.9 &  1.4&  St &   117 &     4.8 &     0.2 &    12 &    $-6$ &    17\\
4 bp004 fr000 &   0.3& $-2.4$ &  0.4 & 0.6&  NSt &    67 &    18.0 &     6.1 &    62 &  $-190$ &    21\\
5 bp006 fr025 &   0.3&  1.0 &  0.7 &  0.6&  St &   126 &    16.4 &     3.9 &    18 &  $-228$ &    22\\
6 bp006 fr059 &   0.02&  6.0 &  0.7 &  0.6&  NSt &   204 &     8.2 &     1.5 &    22 &  $-140$ &    31\\
7 bp007 fr046 &   0.05&  3.1 &  0.8 &  0.7&  NSt &   258 &    18.6 &     6.0 &    78 &  $120$ &    18\\
8 bp007 fr070 &   0.0& $-2.0$ &  0.5 &  0.4&  NSt &   221 &    11.5 &     2.7 &    43 &   $-52$ &    19\\
9 bp007 fr103 &   1.2&  0.8 &  0.6 &  0.9&  St &   205 &    20.6 &     4.9 &     4 &  $-$237 &    17\\
10 bp008 fr021 &   0.2& $-2.7$ &  0.0 &  -- &  NSt &   173 &     6.6 &     1.2 &    86 &   $-$33 &    36\\
11 bp008 fr042 &   0.1& $-2.2$ &  0.6 &  0.7&  St &   220 &     8.5 &     1.3 &     3 &   $-$20 &    16\\
12 bp008 fr056 &   0.4& $-0.6$ &  0.8 &  1.4&  St &   254 &    13.6 &     3.0 &    65 &    $-$2 &    15\\
13 bp009 fr111 &   0.1& $-2.1$ &  0.8 &  0.6&  St &   160 &    24.5 &     7.8 &   199 &   $-$84 &    18\\
14 bp009 fr146 &   0.01&  3.9 &  0.5 &  0.4&  NSt &   175 &    14.5 &     4.0 &   108 &   $-$58 &    18\\
15 bp013 fr075 &   0.1&  0.3 &  0.8 &  1.7&  St &   321 &     7.2 &     1.1 &    52 &   $-$16 &    23\\
16 bp013 fr084 &   1.1&  1.2 &  0.3 &  0.8&  St &   338 &    13.8 &     0.9 &    71 &   $-$78 &    25\\
17 bp013 fr095 &   0.8& $-2.6$ &  1.0 &  0.5&  St &   320 &    16.5 &     1.6 &    16 &   $-$81 &    21\\
18 bp014 fr016 &   0.1&  1.8 &  0.7 &  1.2&  St &   130 &    10.8 &     2.7 &    71 &   $-$75 &    26\\
19 bp014 fr179 &   0.01& $-0.8$ &  0.8 &  1.4&  NSt &   204 &     6.3 &     1.0 &     5 &  $-$114 &    35\\
20 bp015 fr070 &   0.1& $-0.1$ &  0.0 & -- &  NSt &   106 &     9.8 &     2.1 &    43 &  $-$146 &    26\\
21 bp015 fr090 &   0.03& $-5.2$ &  1.0 & 1.1&  NSt &    87 &     5.5 &     1.0 &    12 &   $-$26 &    18\\
22 bp015 fr128 &   0.01&  2.2 &  1.0 &  0.7&  NSt &   107 &     6.5 &     1.2 &    22 &   $-$27 &    19\\
23 bp015 fr133 &   0.01&  2.2 &  1.0 &  0.9&  NSt &   120 &     6.4 &     1.0 &    26 &   $-$61 &    24\\
24 bp016 fr027 &   2.1&  1.5 &  0.8 &  0.8&  St &   121 &    13.2 &     2.1 &    52 &   $-$52 &    22\\
25 bp017 fr177 &   0.01& $-0.2$ &  0.9 &  0.8&  St &   357 &     9.7 &     0.9 &    45 &   $-$34 &    30\\
26 bp017 fr178 &   0.0&  0.2 &  1.0 &  0.8 & St &   385 &     9.9 &     0.9 &    47 &   $-$47 &    31\\
27 bp017 fr229 &   1.9& $-4.7$ & 0.7 &  0.6&  St &   101 &    23.8 &     6.9 &   176 &   $-$66 &    23\\
28 bp017 fr272 &   0.0& $-3.5$ &  0.0 &  -- &  St &   292 &     1.4 &     0.01 &    12 &   $-$12 &    66\\
29 bp017 fr326 &   0.03& $-5.5$ &  0.9 &  0.7&  St &    89 &     7.5 &     1.0 &    30 &    $-$1 &    14\\
30 bp019 fr143 &   1.9& $-0.5$ &  0.9 &  2.7&  St &    95 &    11.2 &     0.8 &    14 &   $-$11 &     8\\
31 bp019 fr154 &   1.0& $-0.9$ & 0.4 &  0.9&  St &   112 &    26.8 &     9.3 &   144 &   $-$64 &    12\\
32 bp022 fr000 &   0.1&  1.8 &  0.6 &  1.4&  St &    67 &    17.0 &     2.6 &    97 &   $-$17 &    12\\
33 bp020 fr032 &   0.1&  4.6 &  0.6 &  0.5&  NSt &   131 &     8.4 &     1.4 &    76 &   $-$61 &    29\\
34 bp020 fr244 &   0.8&  5.8 &  1.0 &  1.2&  St &   118 &    16.0 &     2.0 &    24 &    $-$6 &    13\\
35 bp020 fr255 &   0.6&  5.7 & 0.3 &  0.5&  St &   109 &    21.6 &     4.4 &    12 &  $-$171 &    18\\
36 bp020 fr273 &   0.02&  2.5 &  1.0 &  0.7&  St &   237 &    19.4 &     5.5 &    27 &  $-$302 &    26\\
37 bp021 fr028 &   0.02& -0.9 &  0.8 &  1.1&  NSt &   161 &    12.6 &     2.7 &    11 &  $-$163 &    20\\
38 bp021 fr113 &   0.2& $-0.3$ & 0.7 &  0.9&  NSt &    80 &    17.5 &     3.4 &   106 &   $-$10 &    12\\
39 bp021 fr115 &  30.2& $-3.1$ &  0.0 &  -- &  St &    78 &    19.0 &     2.4 &    11 &   $-$35 &    11\\
40 bp021 fr171 &   0.2& $-2.9$ & 0.2 &  0.6&  St &   115 &     8.3 &     1.3 &    20 &    $-$10 &    13\\
41 bp021 fr217 &   0.1& $-3.4$ &  1.0 &  0.9&  St &   236 &     6.5 &     1.2 &     5 &  $-$110 &    27\\
42 bp022 fr052 &  15.8&  3.9 &  0.0 &  -- &  St &   241 &    33.4 &     8.6 &    96 &  $-$504 &    25\\
43 bp024 fr011 &   0.1&  4.5 & 1.0 &  0.6&  St &    63 &     9.0 &     1.5 &    32 &    $-$1 &    16\\
44 bp024 fr013 &   0.1&  3.4 &  0.1 & 0.5&  St &    75 &    10.7 &     1.2 &    36 &   $-$14 &    17\\
45 bp024 fr014 &   0.2& $-0.5$ &  1.0 &  1.4&  St &    75 &    13.6 &     3.9 &   115 &   $-$83 &    31\\
46 bp024 fr020 &   0.1&  4.4 &  1.0 &  0.6&  St &   207 &    10.2 &     1.4 &    79 &   $-$14 &    28\\
47 bp024 fr021 &   0.1& $-2.6$ & 1.0 &  1.1&  St &   224 &    17.5 &     5.3 &   191 &  $-$220 &    40\\
48 bp024 fr136 &   2.2&  0.3 &  0.4 &  1.0&  NSt &   137 &    17.9 &     5.3 &   109 &   $-$28 &    14\\
49 bp024 fr155 &   0.1&  5.9 &  1.0 &  0.7&  St &   334 &    16.4 &     4.6 &   111 &  $-$103 &    36\\
50 bp024 fr156 &   0.1&  3.9 &  0.0 &  -- &  St &   180 &     8.3 &     0.7 &     6 &   $-$30 &    31\\
51 bp024 fr157 &   0.1&  4.3 &  1.0 &  1.3&  St &   276 &    10.1 &     2.6 &   146 &  $-$136 &    50\\
52 bp024 fr159 &   0.2&  3.1 &  0.5 &  0.7&  St &   302 &    16.8 &     5.3 &    53 &   $-$49 &    27\\
53 bp024 fr160 &   0.3&  3.9 &  0.5 &  0.8&  St &   272 &    18.1 &     5.6 &    47 &   $-$43 &    26\\
54 bp024 fr161 &   1.3&  2.8 &  0.6 &  2.4&  St &   237 &    22.7 &     7.2 &    44 &   $-$93 &    23\\
55 bp024 fr199 &   0.1&  3.5 & 0.6 & 0.4&  St &   430 &    21.4 &     6.3 &    97 &  $-$107 &    31\\
56 bp024 fr202 &   0.8&  6.0 &  1.0 &  0.8&  St &   322 &    17.6 &     3.4 &   313 &   $-$95 &    44\\
57 bp024 fr260 &   1.2&  1.3 &  0.9 &  0.4&  NSt &   251 &    15.7 &     3.0 &    49 &   $-$64 &    28\\
58 bp024 fr263 &  11.7&  1.9 &  0.2 &  1.7&  St &   127 &    26.9 &     4.3 &   173 &   $-$86 &    22\\
59 bp024 fr273 &   0.1&  2.4 &  1.0 &  0.6&  NSt &   307 &    24.0 &     6.7 &   130 &  $-$157 &    27\\
60 bp024 fr026 &   0.4&  2.8 &  0.0 &  --&  St &   286 &    12.8 &     1.2 &    89 &   $-$22 &    26\\
61 bp025 fr029 &   0.5&  1.2 &  0.1 &  0.7&  St &   345 &    21.7 &     6.4 &   259 &  $-$105 &    29\\
62 bp025 fr033 &   0.04&  1.2 & 1.0 &  0.6&  St &   380 &    13.2 &     1.4 &   113 &   $-$73 &    27\\
63 bp025 fr034 &   0.8&  2.0 & 1.0 &  1.4&  St &   359 &    15.1 &     1.8 &   118 &   $-$82 &    24\\
64 bp025 fr092 &   0.4&  3.6 &  0.9 &  0.7&  St &   486 &    12.7 &     1.4 &    26 &   $-$28 &    28\\
65 bp025 fr169 &   0.1&  5.9 &  0.5 &  0.8&  St &   431 &    12.4 &     1.5 &     6 &   $-$17 &    26\\
66 bp025 fr197 &   0.8&  4.1 &  0.6 &  0.8&  St &   453 &    18.6 &     4.3 &   214 &   $-$93 &    34\\
67 bp026 fr181 &  38.8& $-2.7$ & 0.7 & 0.7&  St &    91 &    45.2 &    13.0 &   129 &  $-$131 &    15\\
68 bp026 fr187 &  59.0& $-3.1$ &  0.3 &  0.8&  NSt &   101 &    35.8 &     7.0 &   240 &   $-$20 &    15\\
69 bp026 fr190 &  66.2& $-2.3$ &  0.7 &  0.7&  St &   100 &    35.9 &     7.8 &    40 &   $-$77 &    13\\
70 bp026 fr202 &  99.0& $-3.0$ &  0.7 &  0.7&  St &   107 &    30.8 &     4.7 &    33 &   $-$16 &    13\\
71 bp026 fr224 &  23.4& $-3.1$ &  1.0 &  0.9&  St &   104 &    44.8 &    12.6 &   141 &  $-$136 &    15\\
72 bp026 fr227 &   0.3& $-2.9$ & 0.0 &  --&  St &   133 &    31.2 &     5.6 &    23 &   $-$91 &    15\\
73 bp026 fr228 &   1.6& $-2.8$ &  0.2 &  2.1&  St &   131 &    25.9 &     1.9 &    24 &   $-$28 &    13\\
74 bp026 fr332 &   0.8& $-2.2$ &  0.7 &  0.6&  NSt &   160 &    32.0 &     8.0 &    47 &    $-$5 &     8\\
75 bp026 fr375 &   0.5&  0.1 &  1.0 &  0.4&  St &   140 &    40.5 &    12.0 &   244 &  $-$359 &    20\\
76 bp027 fr143 &   0.8&  2.2 &  1.0 &  0.8&  St &   106 &    12.9 &     1.6 &     4&   $-$94 &    19\\
77 bp027 fr170 &   0.1& 0.0 &  1.0 &  8.6&  St &   142 &     8.3 &     1.6 &     4 &   $-$12 &    12\\
78 bp029 fr031 &   0.1&  4.4 &  0.9 & 5.2&  St &   106 &     7.1 &     1.3 &     4 &    $-$2 &    13\\
79 bp029 fr091 &   0.1&  3.5 &  0.0 &  --&  St &   114 &    11.4 &     0.5 &    39 &   $-$12 &    20\\
80 bp029 fr252 &   0.0&  1.9 &  0.7 &  0.9&  St &   188 &     7.1 &     1.2 &    70 &   $-$14 &    23\\
81 bp029 fr264 &   0.0&  1.5 &  0.7 &  0.7&  St &   138 &     5.2 &     0.7 &    13 &   $-$20 &    21\\
82 bp030 fr039 &   0.6& $-4.9$ &  0.0 &  --&  NSt &   173 &    17.1 &     4.8 &   198 &   $-$28 &    25\\
83 bp030 fr115 &   0.04&  1.8 &  1.0 &  0.6&  St &   144 &     9.0 &     1.2 &    26 &  $-$147 &    33\\
84 bp032 fr355 &  10.3& $-5.6$ &  0.0 & --&  NSt &    69 &    17.4 &     1.5 &    65 &    $-$0.3 &    13\\
85 bp033 fr157 & 278.9&  1.2 &  0.0 &  --&  St &    58 &    60.5 &    12.7 &    69 &    $-$9 &     5\\
86 bp033 fr122 &   0.1& $-1.0$ &  0.9 &  0.6&  NSt &   104 &     6.3 &     1.6 &    56 &  $-$112 &    31\\
87 bp038 fr424 &  99.5& $-2.5$ &  0.6 &  0.5&  St &    75 &    40.1 &     9.9 &    60 &   $-$13 &     6\\
88 bp039 fr012 &  59.6& $-3.2$ &  0.5 &  0.6&  St &    94 &    24.1 &     7.0 &    73 &  $-$117 &    13\\
89 bp039 fr102 &   3.7& $-2.2$ &  0.5 &  0.8&  St &   111 &    23.9 &     6.2 &    26 &  $-$150 &    14\\
90 bp039 fr110 &   0.7& $-0.8$ &  0.4 &  1.2&  St &   116 &    17.6 &     3.6 &     4&   $-$70 &    13\\
91 bp039 fr146 &   0.03&  2.4 &  0.5 &  0.5&  NSt &   161 &    21.7 &     6.2 &    43 &   $-$87 &    12\\
92 bp039 fr203 &   0.3&  2.2 &  1.0 & 0.4&  St &   124 &    16.8 &     3.0 &    96 &   $-$93 &    18\\
93 bp039 fr212 &   0.2&  0.1 &  1.0 & 8.6 &  St &   129 &    14.9 &     3.1 &    92 &   $-$81 &    16\\
94 bp039 fr262 &  25.3& -0.1 &  1.0 &  0.5&  St &   102 &    30.8 &    11.3 &   219 &   $-$27 &    14\\
95 bp040 fr015 &   0.2&  0.5 &  1.0 &  0.7&  St &    74 &    16.0 &     5.0 &   164 &  $-$185 &    35\\
96 bp040 fr022 &   0.1& $-1.7$ &  1.0 &  0.7&  St &   224 &    16.6 &     4.8 &   234 &  $-$203 &    41\\
97 bp040 fr027 &   0.02&  0.1 &  1.0 &  0.8&  St &   359 &    10.7 &     2.0 &   170 &   $-$28 &    33\\
98 bp040 fr038 &   0.1& -0.5 &  0.7 & 0.7 &  St &   315 &    14.6 &     3.2 &   120 &   $-$70 &    30\\
99 bp041 fr036 &   0.5& -2.6 &  0.1 & 0.8&  St &   152 &    20.5 &     4.5 &     3 &  $-$121 &    13\\
100 bp042 fr067 &   0.1&  2.0 &  0.2 & 6.2&  NSt &   237 &     8.4 &     2.3 &    99 &  $-$141 &    39\\
101 bp043 fr012 &   0.3&  1.8 &  0.1 & 0.6&  St &   195 &     9.8 &     1.6 &    90 &   $-$50 &    26\\
102 bp043 fr018 &   0.1&  1.7 &  0.9 &  0.5&  St &   182 &     8.4 &     0.5 &    25 &   $-$23 &    20\\
103 bp043 fr163 &   0.02&  4.8 &  1.0 &  0.7&  St &   125 &     4.2 &     0.6 &    46 &   $-$11 &    30\\
104 bp044 fr073 &   7.2& $-1.3$ &  1.0 &  3.9&  NSt &   133 &    35.2 &    13.3 &   117 &   $-$72 &    12\\
105 bp044 fr088 &  58.3& $-0.9$ &  0.9 & 0.6&  St &    94 &    20.4 &     2.5 &     9 &   $-$14 &    11\\
106 bp044 fr157 &   0.6&  1.9 &  1.0 &  2.0&  St &   229 &    19.1 &     6.9 &   194 &   $-$47 &    25\\
107 bp044 fr159 &   0.2& $-0.7$ &  1.0 &  1.7&  St &   237 &    20.0 &     7.3 &   220 &   $-$44 &    25\\
108 bp044 fr160 &   0.01& $-3.3$ &  0.6 &  1.8&  NSt &   303 &     7.0 &     1.9 &   173 &   $-$55 &    48\\
109 bp045 fr094 &   0.2&  3.0 &  0.5 &  0.4&  NSt &   101 &    18.9 &     5.1 &   158 &   $-$38 &    15\\
110 bp046 fr038 &  15.9&  0.1 &  0.3 &  0.5&  St &   218 &    27.3 &     7.6 &   200 &   $-$96 &    18\\
111 bp046 fr040 &   0.00& $-0.5$ &  0.3 & 0.8&  St &   204 &     5.0 &     0.6 &    22 &   $-$22 &    37\\
112 bp046 fr052 &   0.9& $-0.6$ &  1.0 &  1.5&  St &   268 &    11.2 &     3.2 &   105 &   $-$72 &    42\\
113 bp046 fr056 &   1.7&  5.5 &  0.5 &  0.6&  St &   320 &    10.5 &     2.1 &    41 &   $-$11 &    32\\
114 bp046 fr066 &   0.03& $-0.8$ &  1.0 & 1.5&  NSt &   593 &    13.5 &     3.5 &   273 &  $-$176 &    58\\
115 bp046 fr205 &   1.4&  2.6 &  0.3 &  1.1&  St &   518 &    30.0 &     9.8 &   341 &  $-$177 &    31\\
116 bp053 fr000 &  46.0& $-5.6$ &  0.5 &  0.9&  St &    81 &    27.8 &     2.0 &    37 &   $-$54 &    11\\
117 bp053 fr025 &  22.7& $-4.2$ &  0.7 &  1.2&  St &    66 &    39.8 &    12.5 &   119 &   $-$81 &     10\\
118 bp053 fr216 &   0.2&  0.42 &  0.8 &  0.6&  St &    69 &    11.3 &     1.0 &    50 &   $-$16 &    13\\
119 bp064 fr022 &   0.02& $-2.8$ &  0.9 &  0.5&  St &   170 &     8.7 &     1.0 &    90&   $-$52 &    38\\
120 bp064 fr025 &   0.00& $-1.5$ &  0.0 &  --&  St &   343 &     7.9 &     2.0 &   394 &   $-$58 &    81\\
121 bp070 fr094 &  26.3& $-2.2$ &  0.4 &  0.5&  St &    72 &    35.1 &    11.2 &   211 &  $-$105 &    14\\
122 bp083 fr006 &   0.8&  1.7 &  0.0 &  --&  St &    91 &    11.8 &     1.2 &    23 &    $-$5 &    12\\
123 bp083 fr029 &   0.3&  5.9 &  0.9 &  0.8&  St &    72 &     3.9 &     0.4 &     8 &    $-$1 &    18\\
124 bp083 fr040 &   0.04&  4.8 &  0.9 &  0.8&  St &   118 &     8.3 &     2.2 &   103 &   $-$29 &    28\\
125 bp085 fr001 &  32.4& $-2.1$ &  0.9 & 0.6&  St &    72 &    40.1 &    11.6 &    27 &   $-$41 &     6\\
126 bp090 fr001 &   0.1& $-3.7$ &  0.8 &  0.6&  NSt &   169 &     4.9 &     1.0 &   120 &   $-$79 &    59\\
\hline
\label{table2}
\end{longtable}
\end{center}

\twocolumn

\section{Additional figures}
\label{Additional_figures}

\begin{figure*}
\vspace{3cm}
\centering
\includegraphics[scale=0.13]{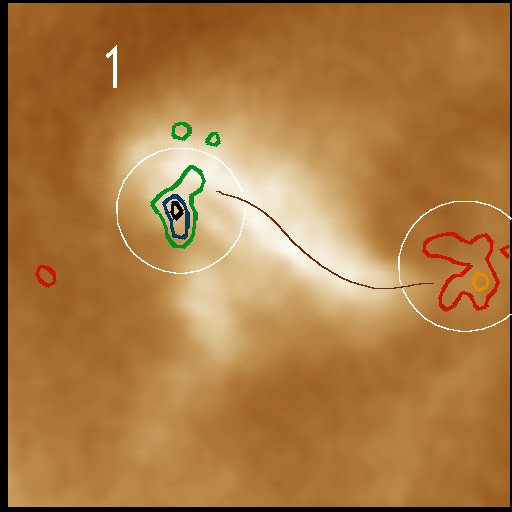}
\includegraphics[scale=0.13]{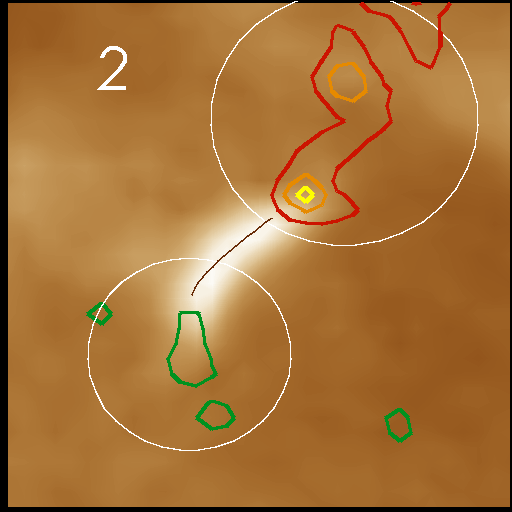}
\includegraphics[scale=0.13]{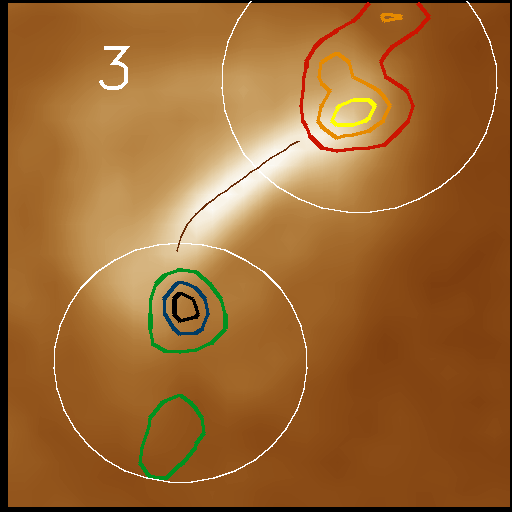}
\includegraphics[scale=0.13]{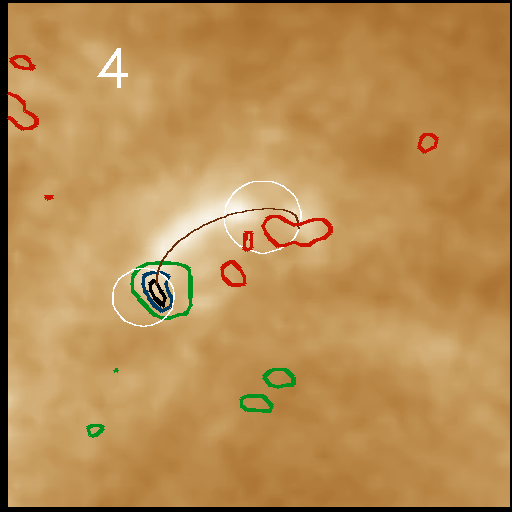}
\includegraphics[scale=0.13]{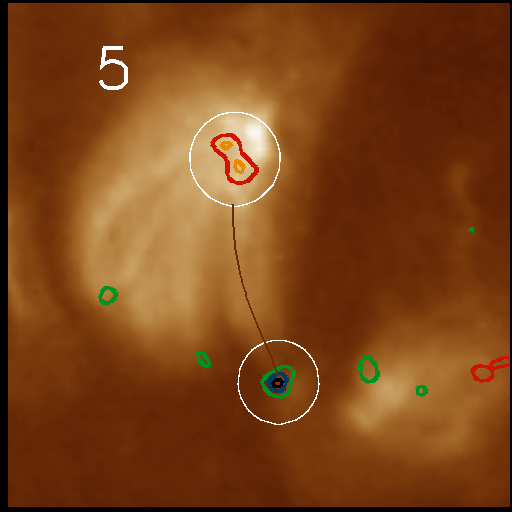}
\includegraphics[scale=0.13]{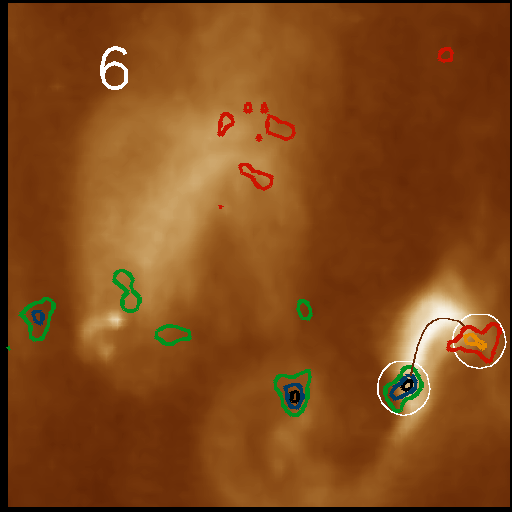}\\
\includegraphics[scale=0.13]{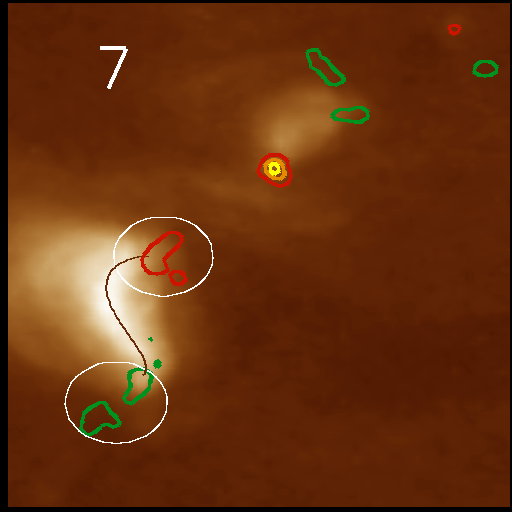}
\includegraphics[scale=0.13]{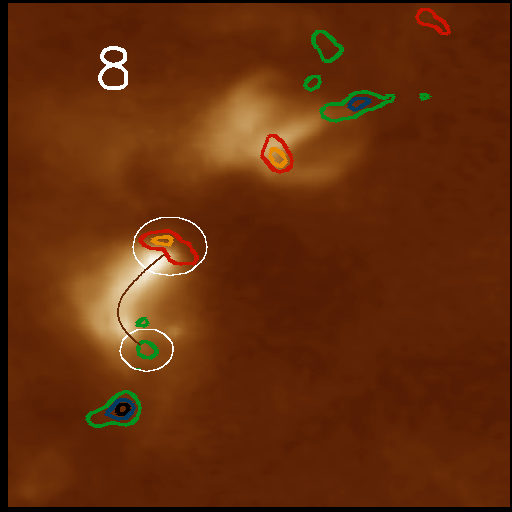}
\includegraphics[scale=0.13]{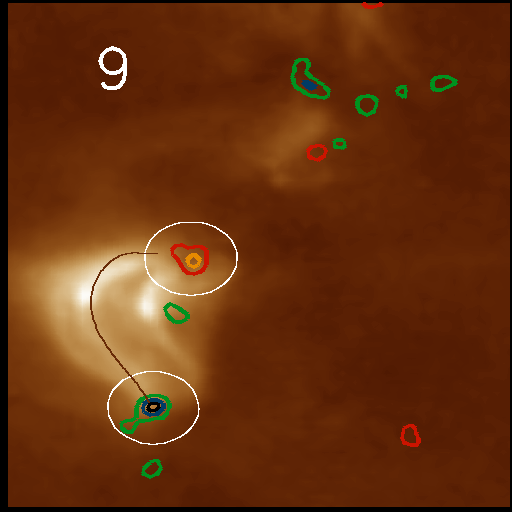}
\includegraphics[scale=0.13]{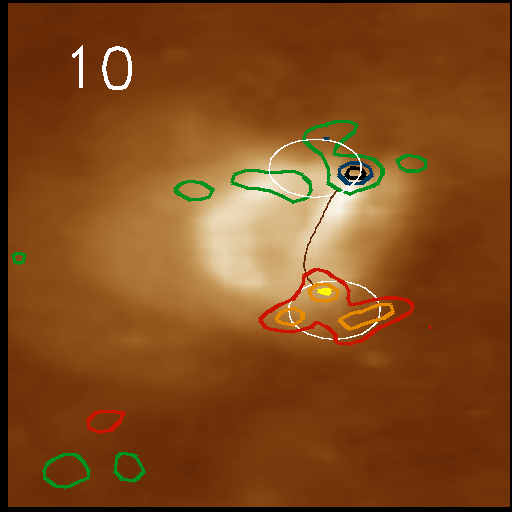}
\includegraphics[scale=0.13]{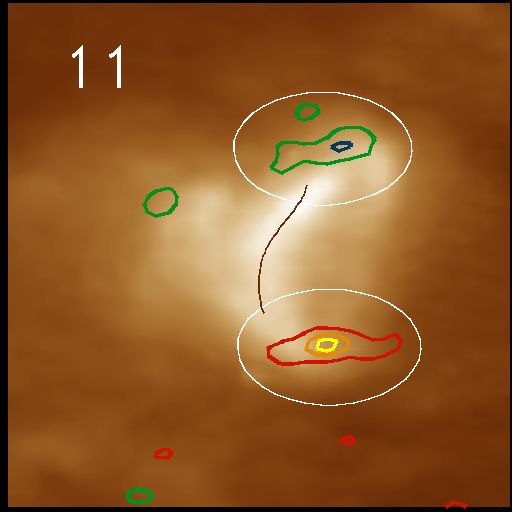}
\includegraphics[scale=0.13]{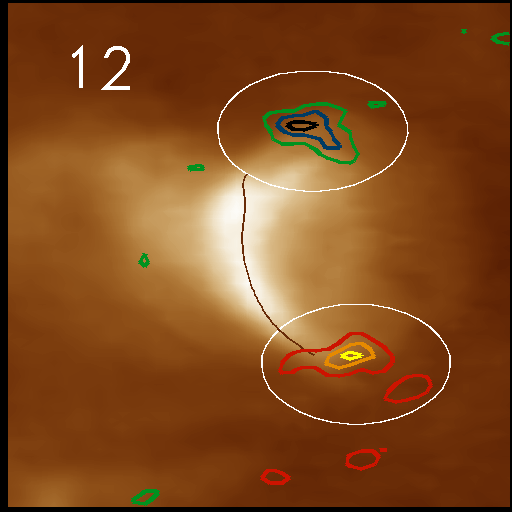}\\
\includegraphics[scale=0.13]{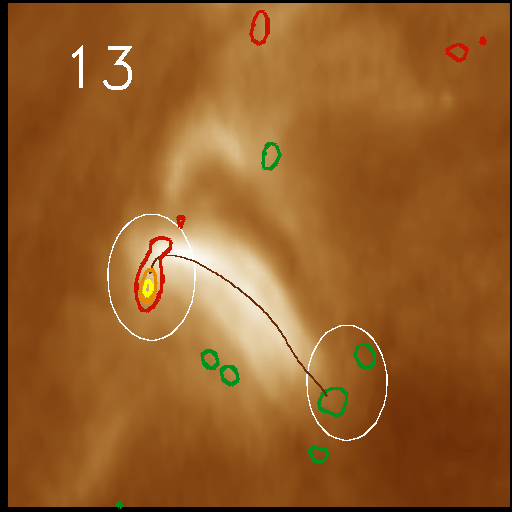}
\includegraphics[scale=0.13]{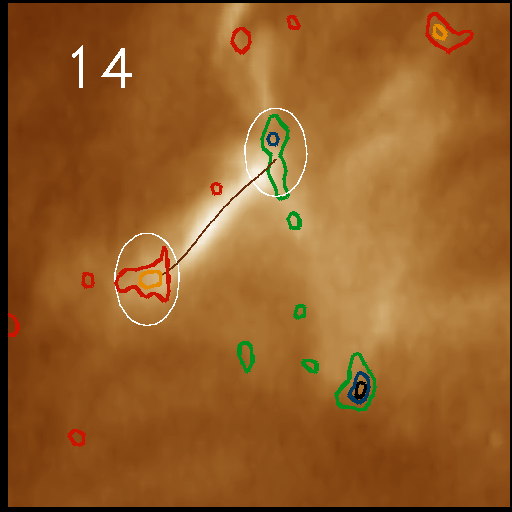}
\includegraphics[scale=0.13]{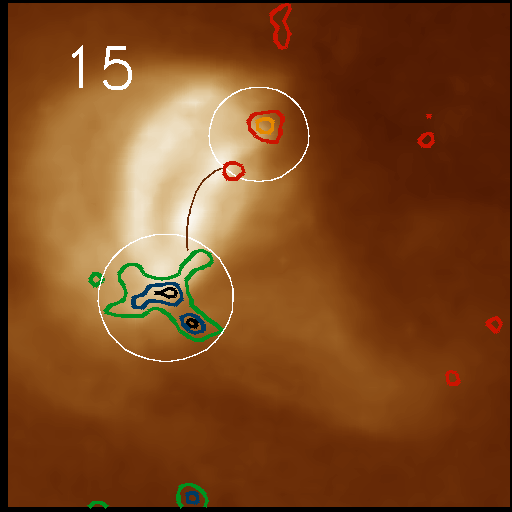}
\includegraphics[scale=0.13]{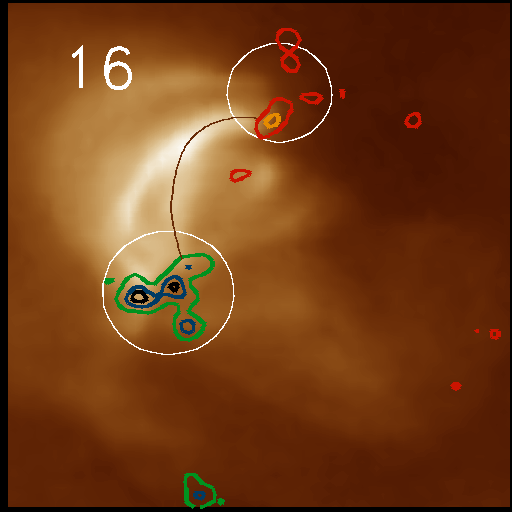}
\includegraphics[scale=0.13]{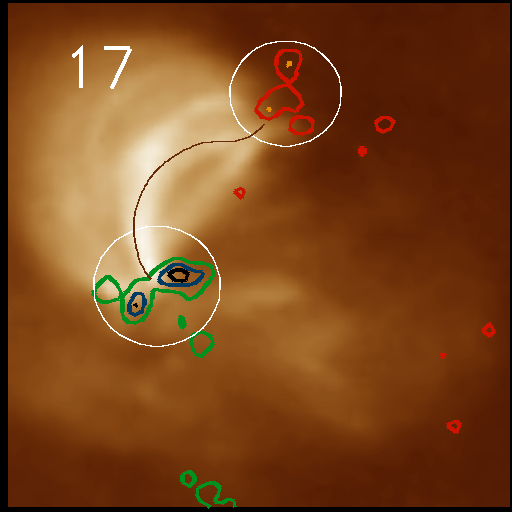}
\includegraphics[scale=0.13]{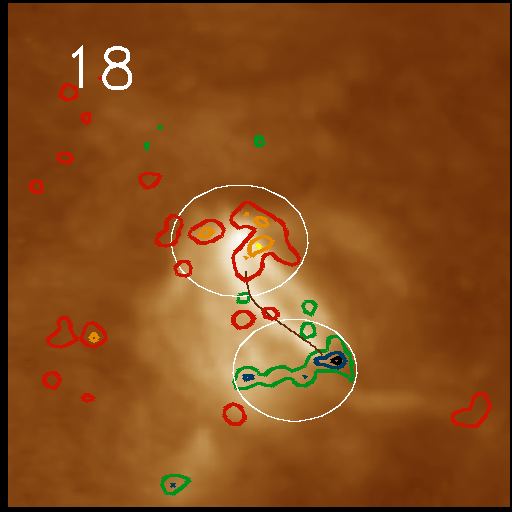}\\
\includegraphics[scale=0.13]{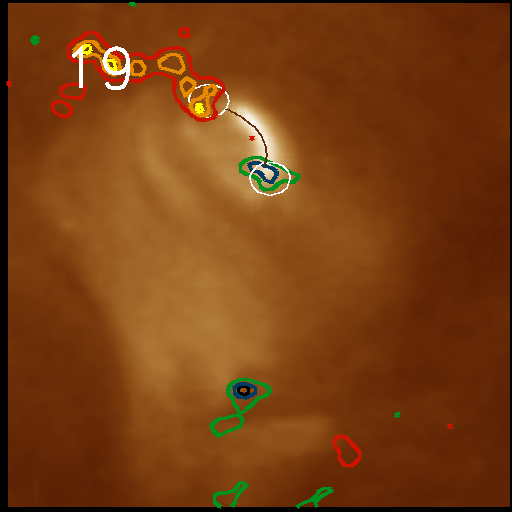}
\includegraphics[scale=0.13]{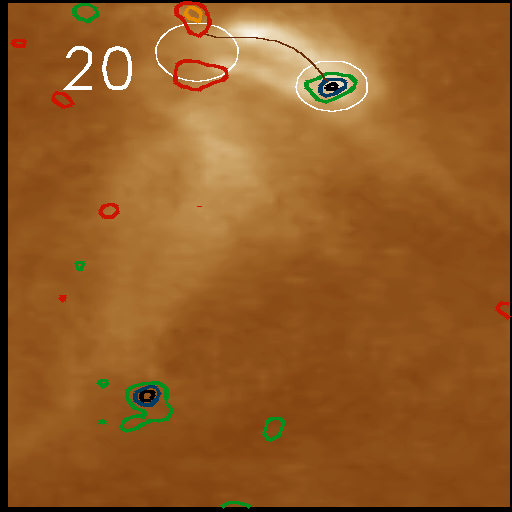}
\includegraphics[scale=0.13]{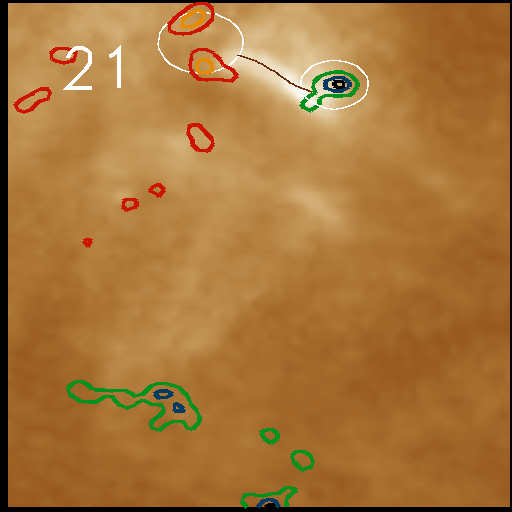}
\includegraphics[scale=0.13]{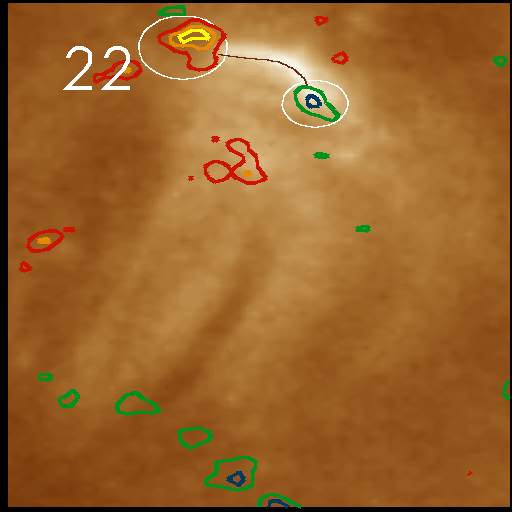}
\includegraphics[scale=0.13]{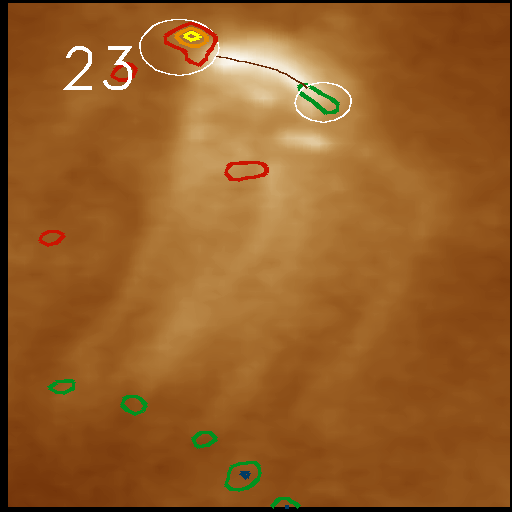}
\includegraphics[scale=0.13]{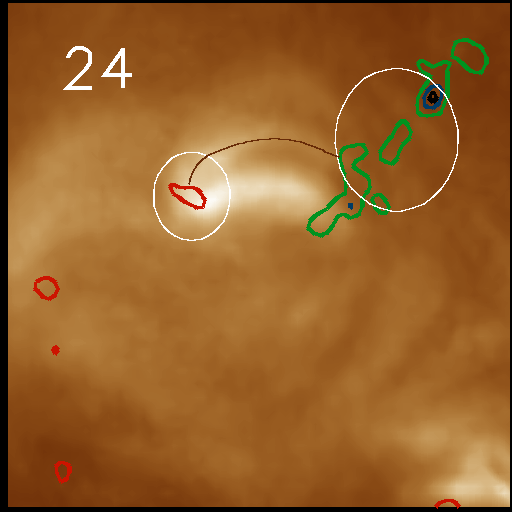}\\
\includegraphics[scale=0.13]{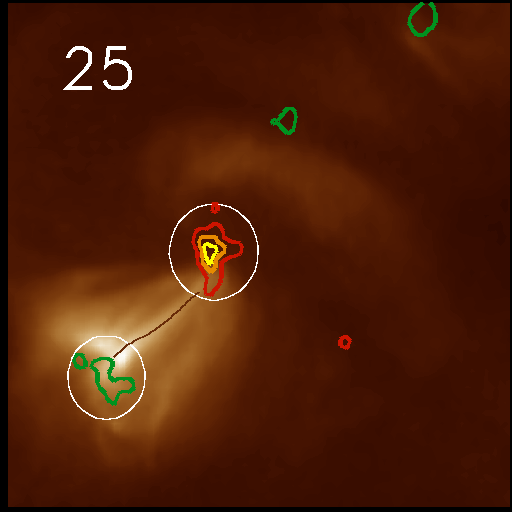}
\includegraphics[scale=0.13]{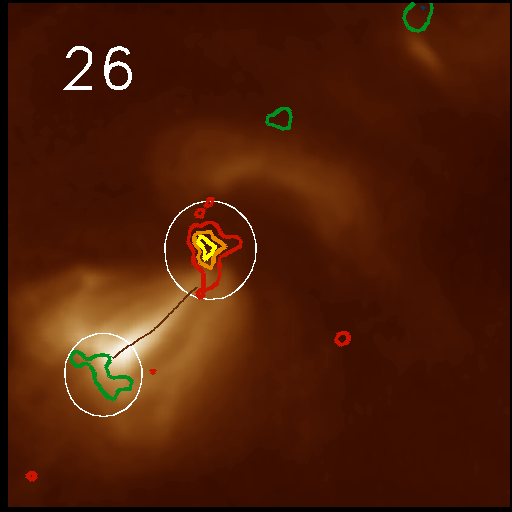}
\includegraphics[scale=0.13]{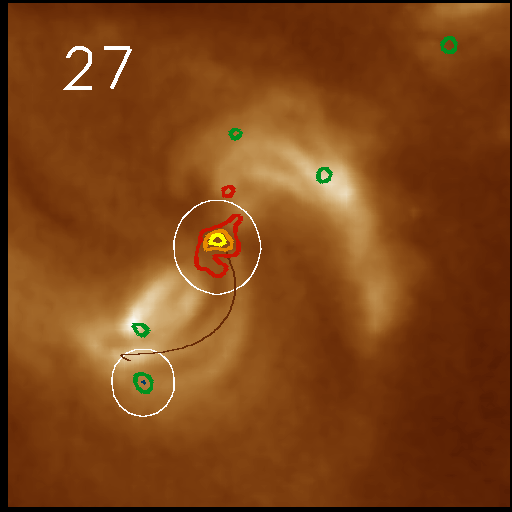}
\includegraphics[scale=0.13]{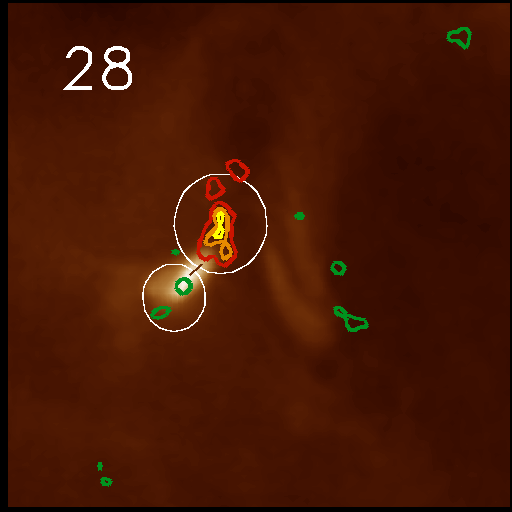}
\includegraphics[scale=0.13]{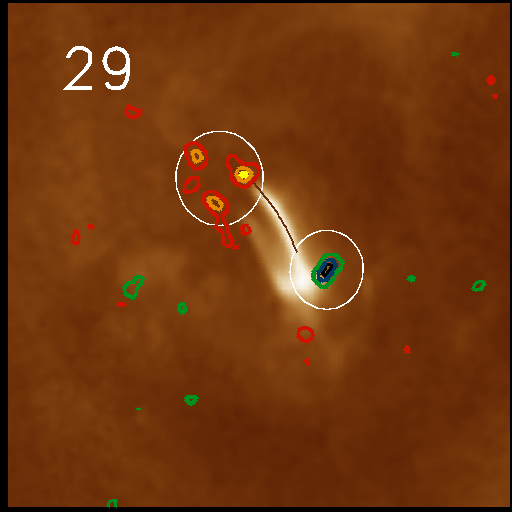}
\includegraphics[scale=0.13]{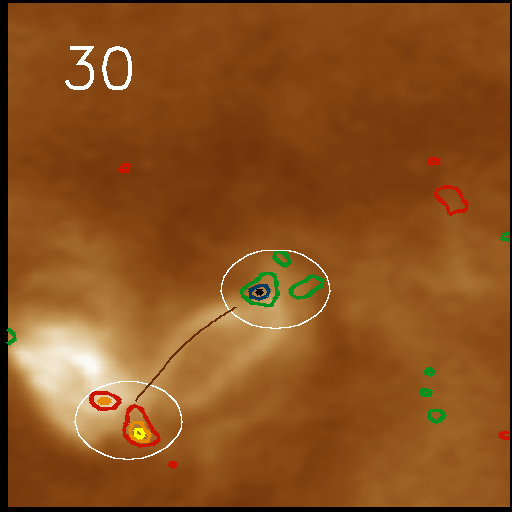}\\
\includegraphics[scale=0.13]{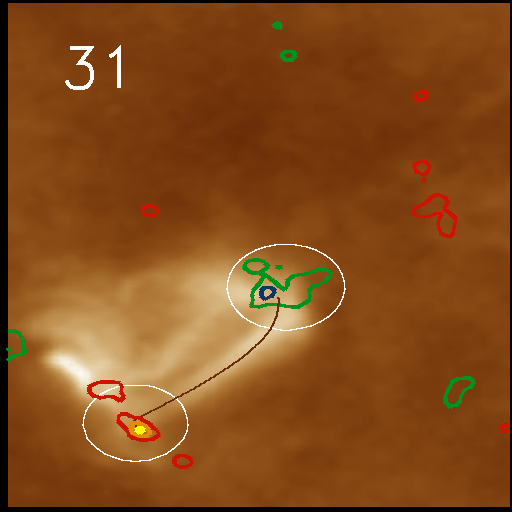}
\includegraphics[scale=0.13]{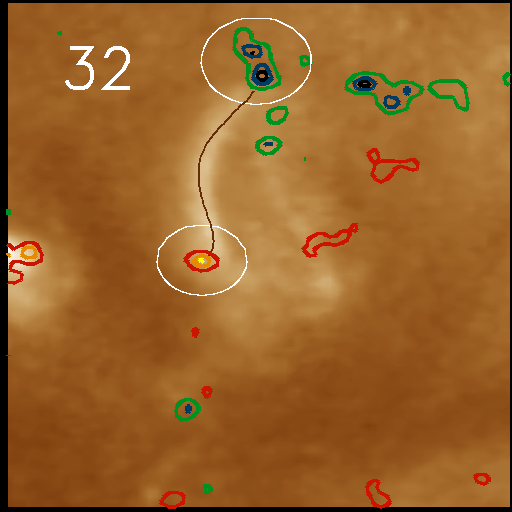}
\includegraphics[scale=0.13]{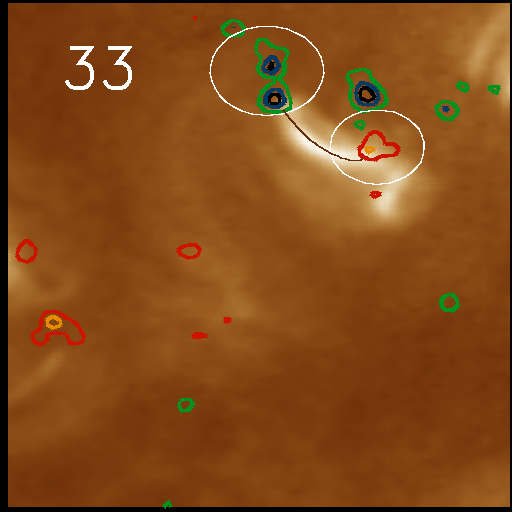}
\includegraphics[scale=0.13]{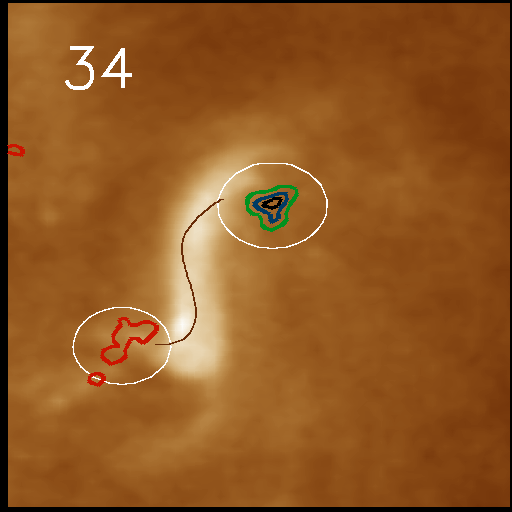}
\includegraphics[scale=0.13]{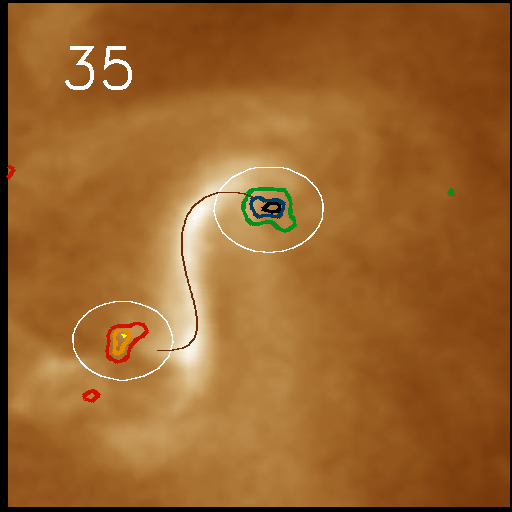}
\includegraphics[scale=0.13]{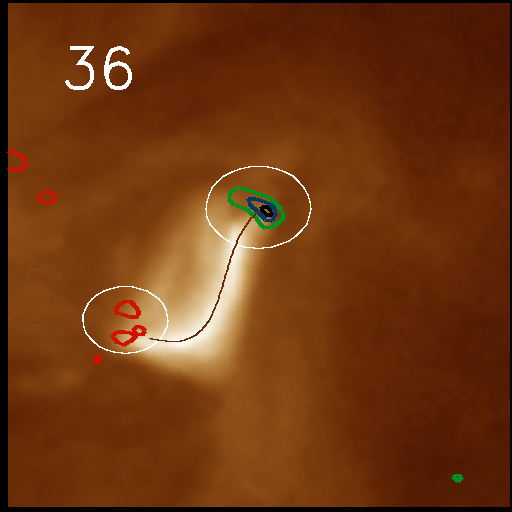}\\
\includegraphics[scale=0.13]{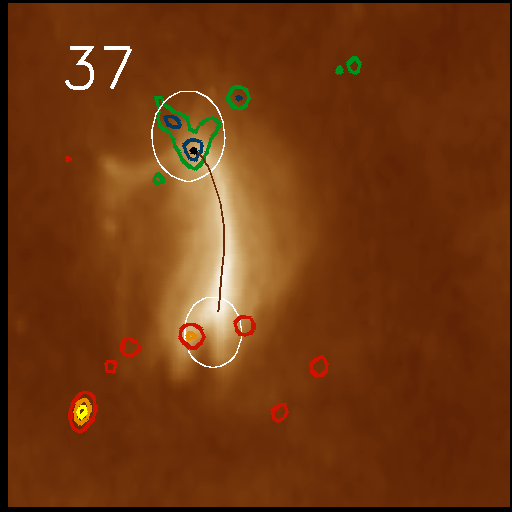}
\includegraphics[scale=0.13]{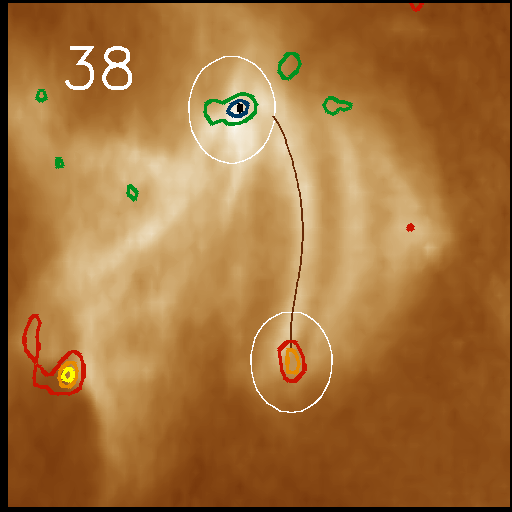}
\includegraphics[scale=0.13]{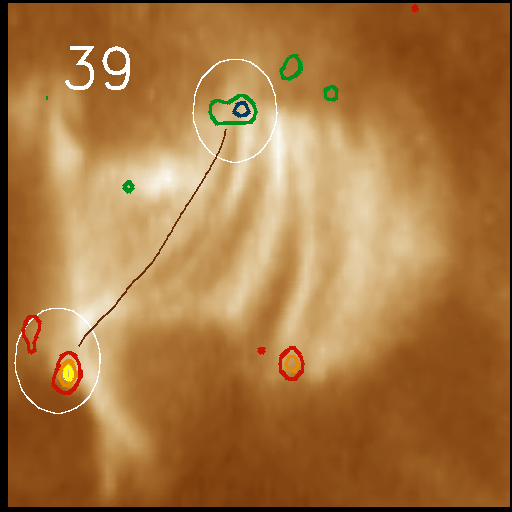}
\includegraphics[scale=0.13]{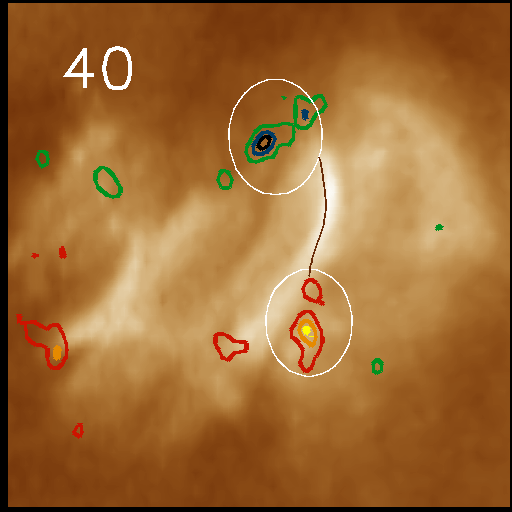}
\includegraphics[scale=0.13]{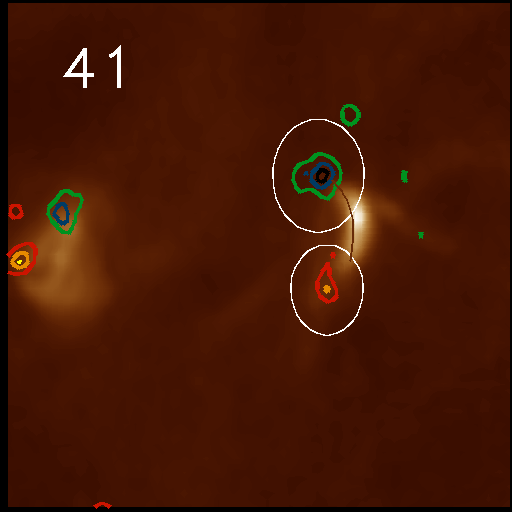}
\includegraphics[scale=0.13]{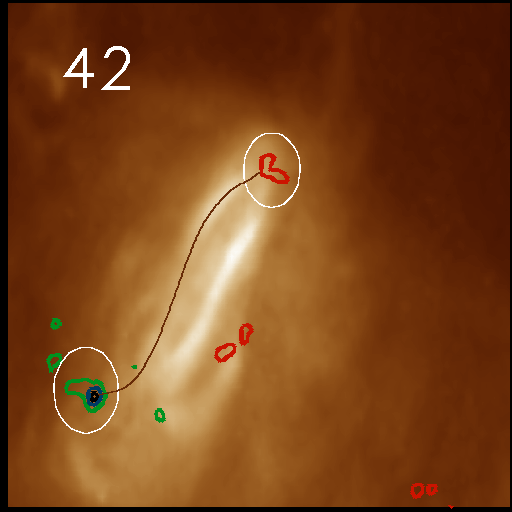}\\
\caption{Loop extrapolation examples (1--42) obtained from the LMHS modelling in the same format  as in the left panels of Fig.~\ref{fig2}.  The black and green contours indicate negative polarities. The orange and red contours outline positive polarities. The extrapolated magnetic field line is shown with a black colour line. All contours are relative to the absolute maximum field strength in the magnetograms, equispaced between 0 and max $|B_z|$. The intensity of each image is scaled to the maximum intensity of the image.}
\label{app-fig1}
\end{figure*}
 
\begin{figure*}
\vspace{3cm}
\centering
\includegraphics[scale=0.13]{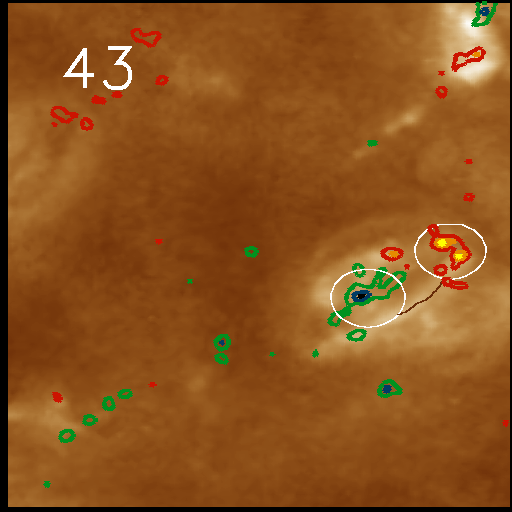}
\includegraphics[scale=0.13]{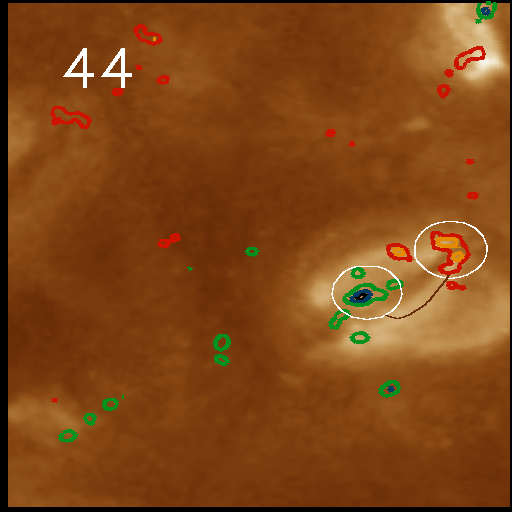}
\includegraphics[scale=0.13]{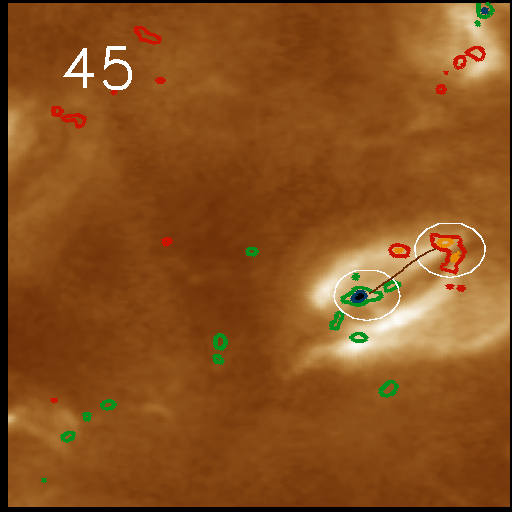}
\includegraphics[scale=0.13]{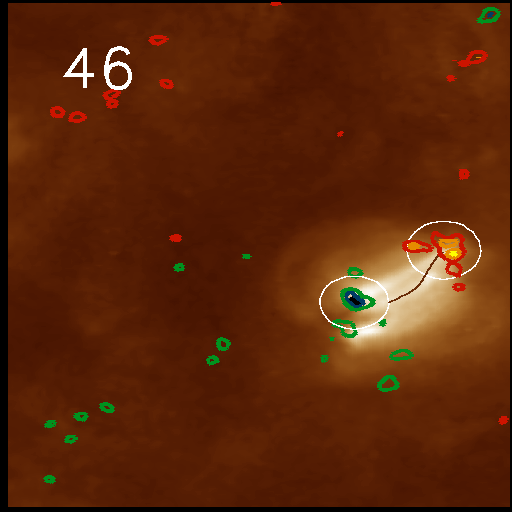}
\includegraphics[scale=0.13]{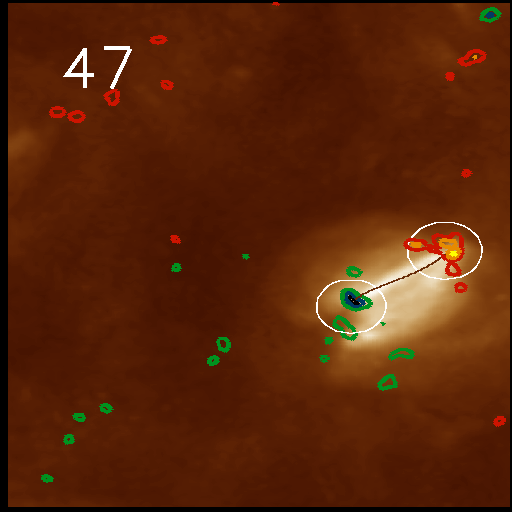}
\includegraphics[scale=0.13]{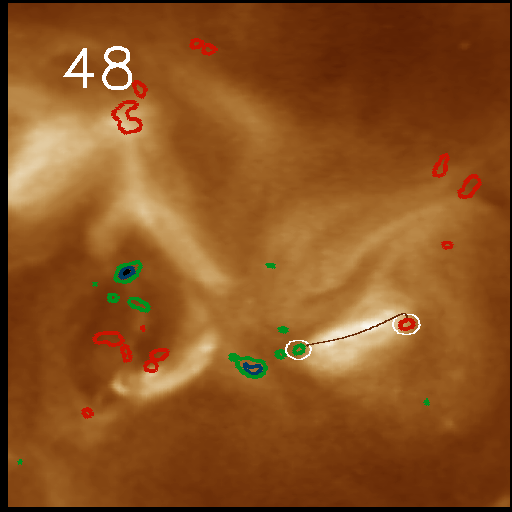}\\
\includegraphics[scale=0.13]{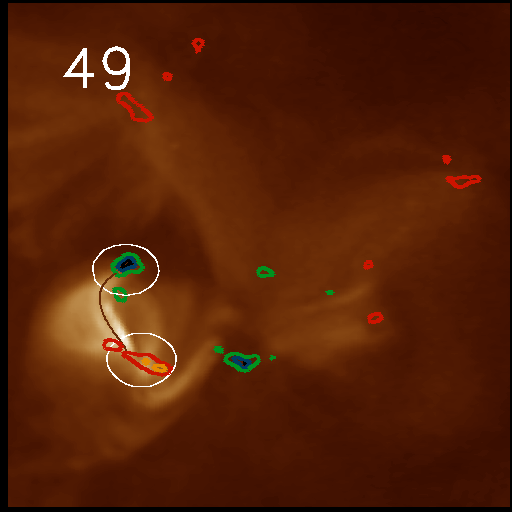}
\includegraphics[scale=0.13]{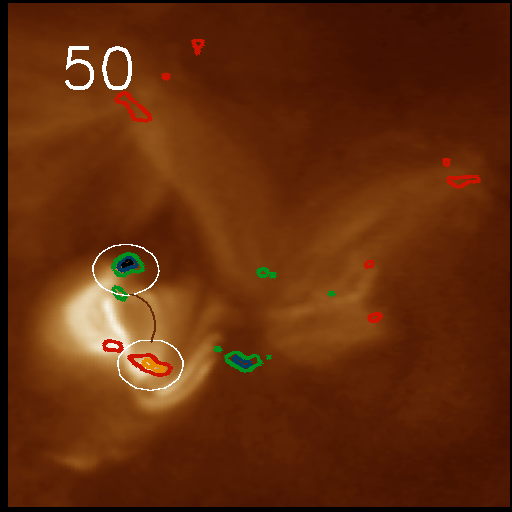}
\includegraphics[scale=0.13]{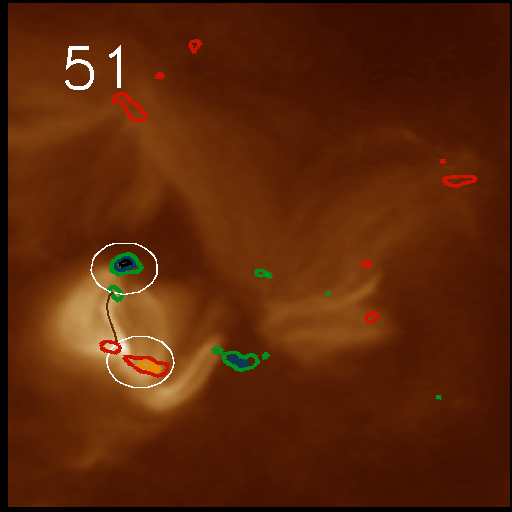}
\includegraphics[scale=0.13]{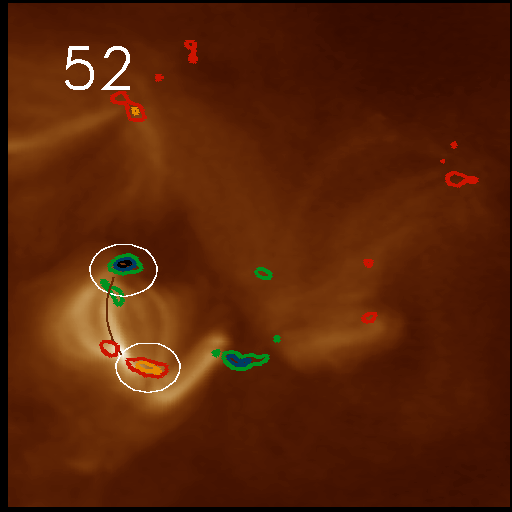}
\includegraphics[scale=0.13]{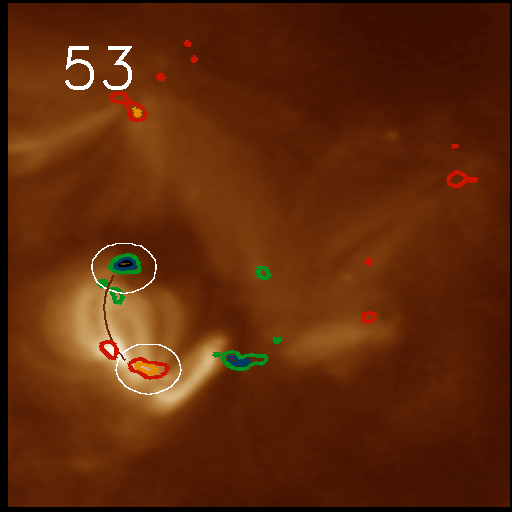}
\includegraphics[scale=0.13]{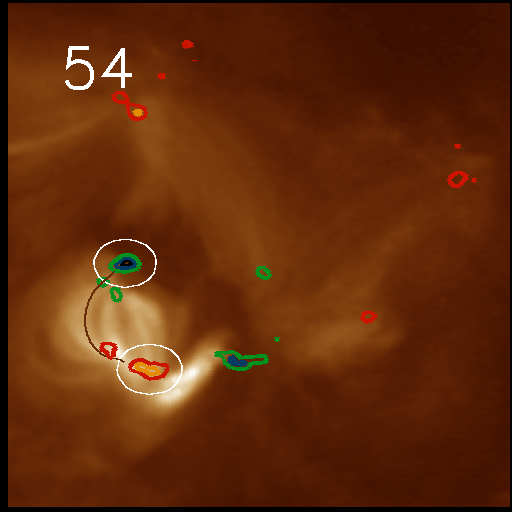}\\
\includegraphics[scale=0.13]{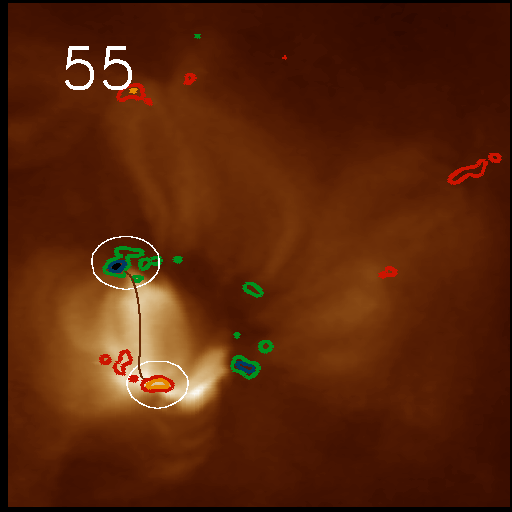}
\includegraphics[scale=0.13]{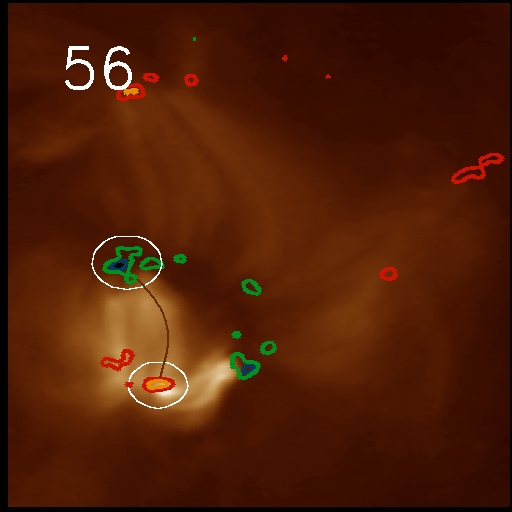}
\includegraphics[scale=0.13]{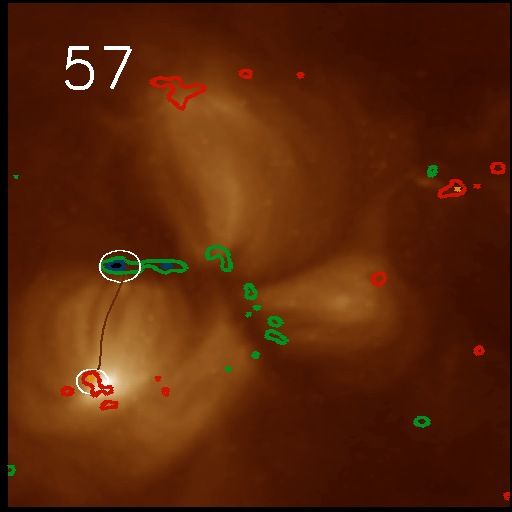}
\includegraphics[scale=0.13]{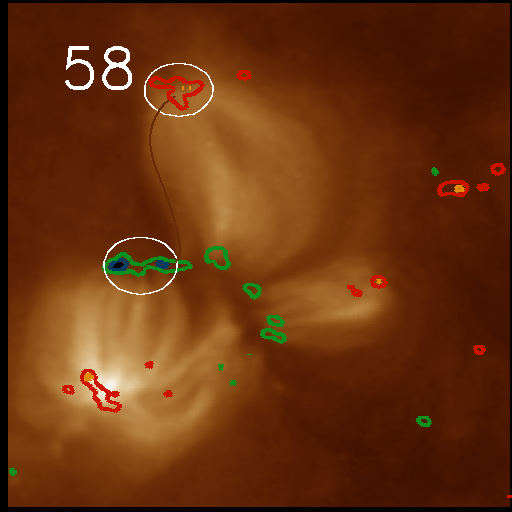}
\includegraphics[scale=0.13]{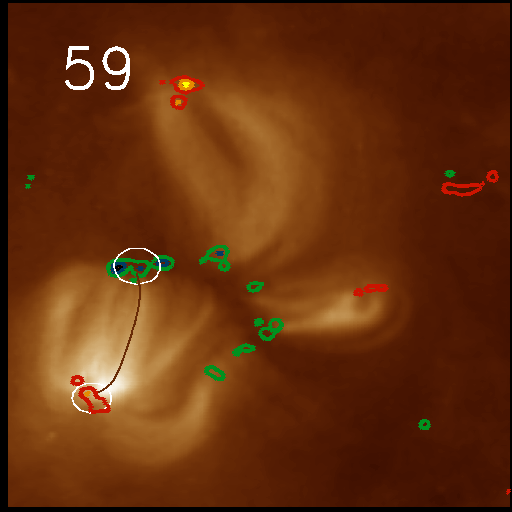}
\includegraphics[scale=0.13]{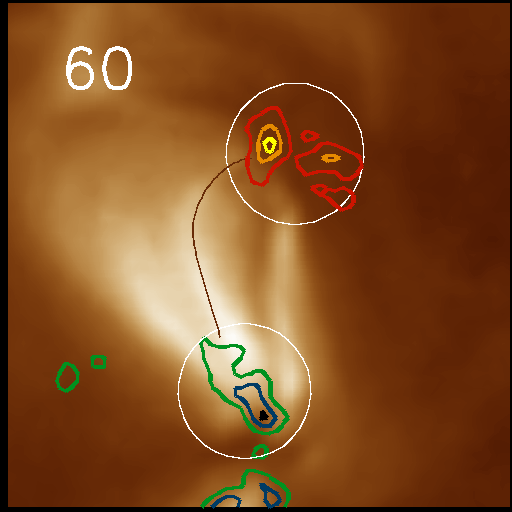}\\ 
\includegraphics[scale=0.13]{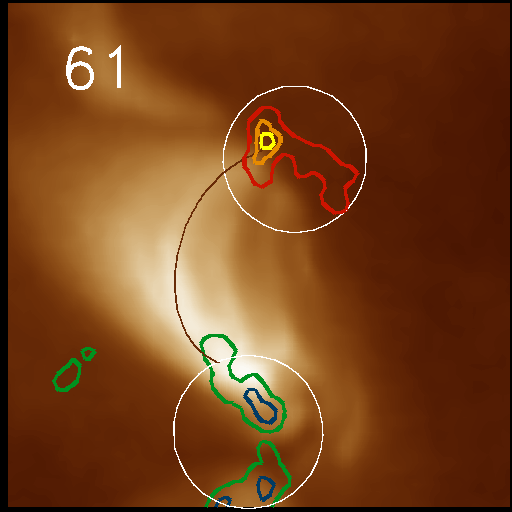}
\includegraphics[scale=0.13]{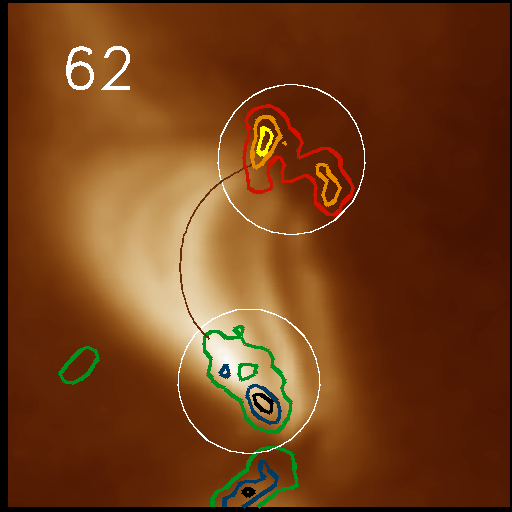}
\includegraphics[scale=0.13]{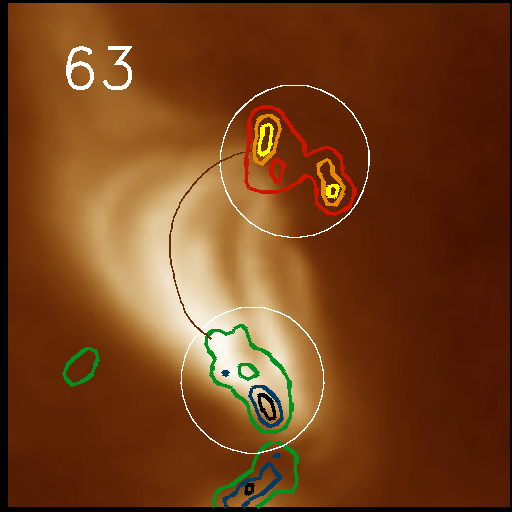}
\includegraphics[scale=0.13]{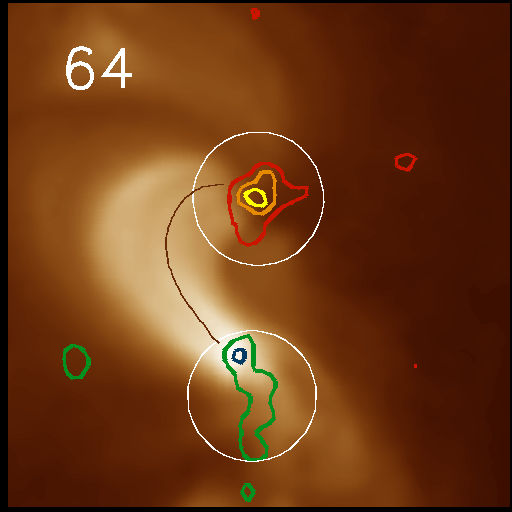}
\includegraphics[scale=0.13]{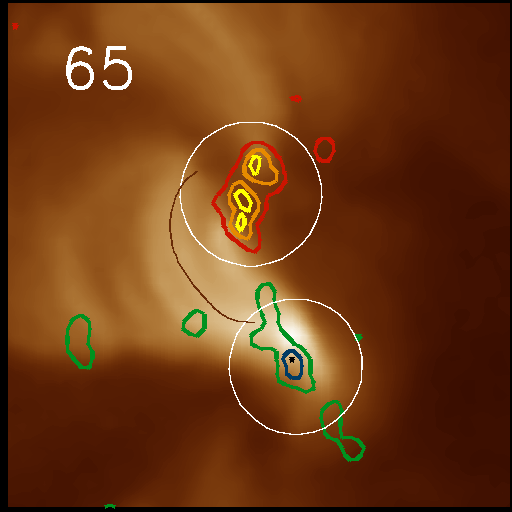}
\includegraphics[scale=0.13]{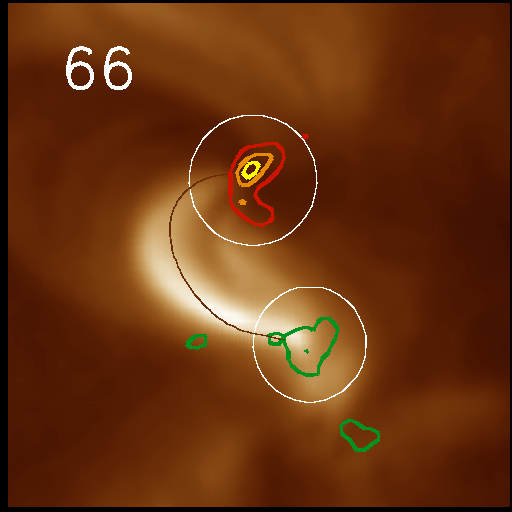}\\ 
\includegraphics[scale=0.13]{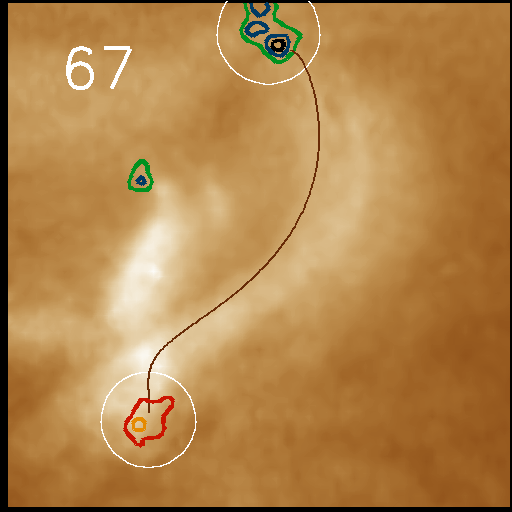}
\includegraphics[scale=0.13]{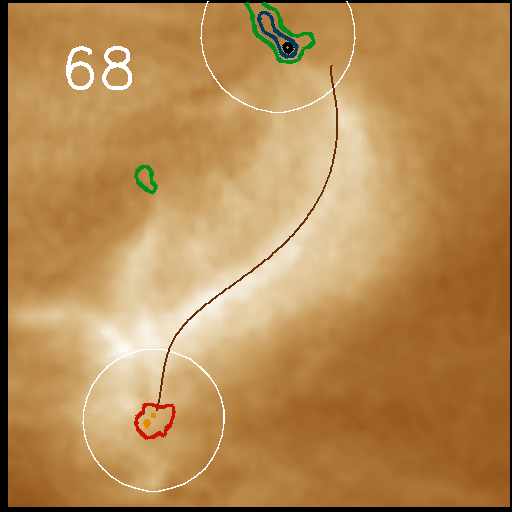}
\includegraphics[scale=0.13]{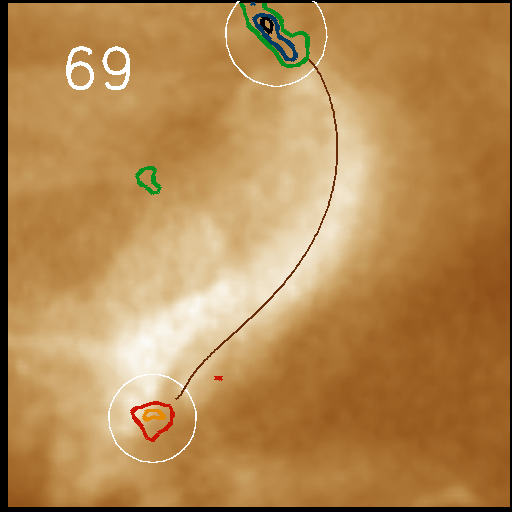}
\includegraphics[scale=0.13]{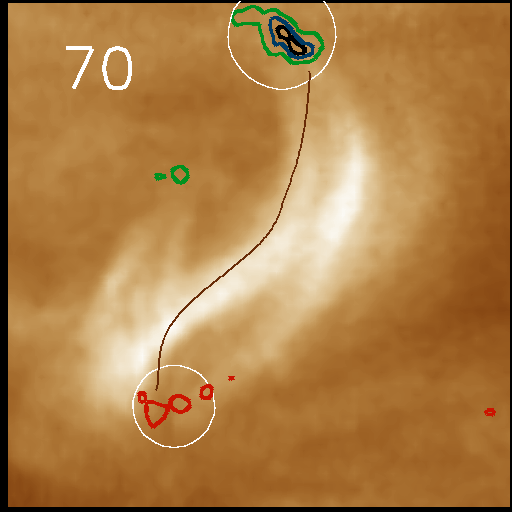}
\includegraphics[scale=0.13]{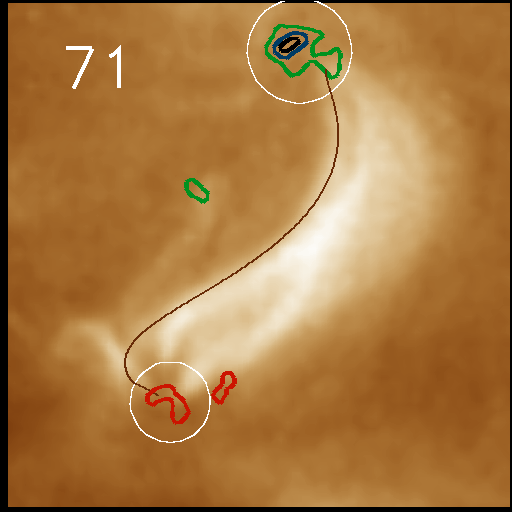}
\includegraphics[scale=0.13]{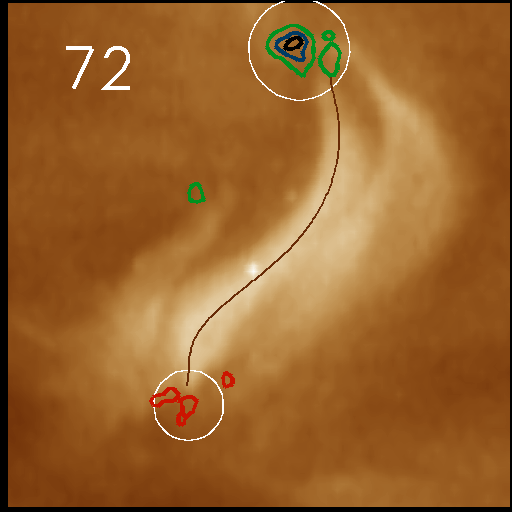}\\ 
\includegraphics[scale=0.13]{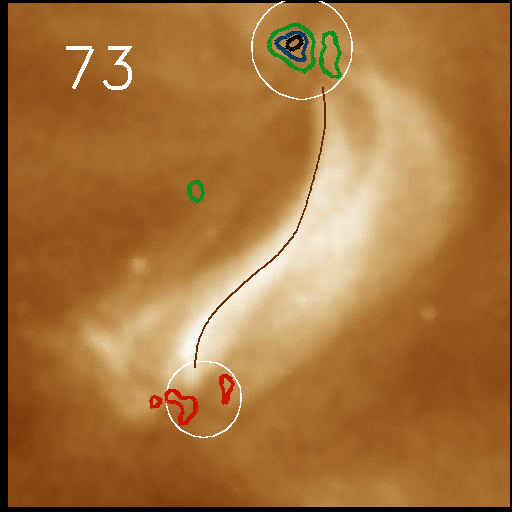}
\includegraphics[scale=0.13]{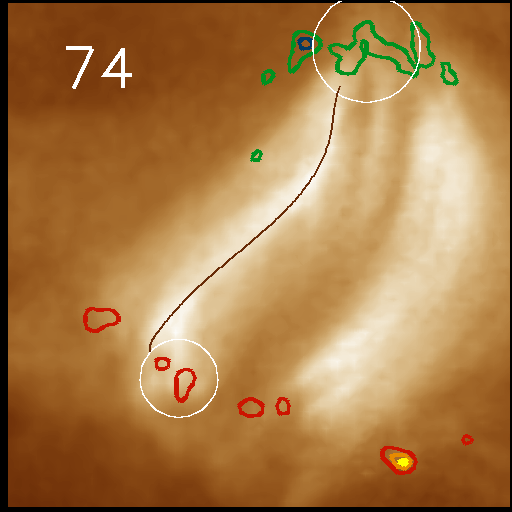}
\includegraphics[scale=0.13]{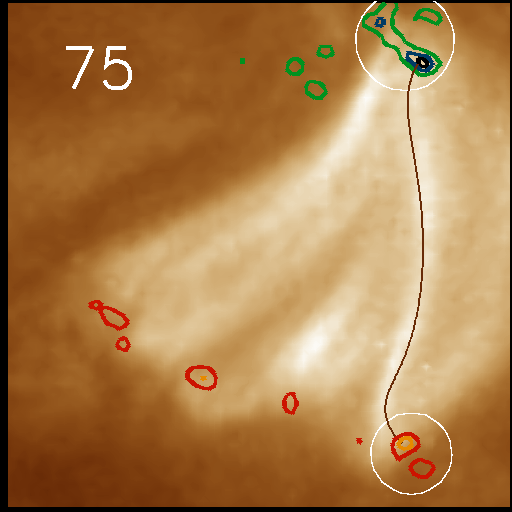}
\includegraphics[scale=0.13]{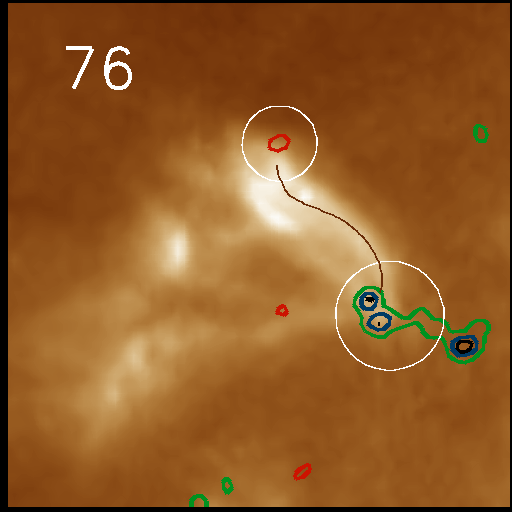}
\includegraphics[scale=0.13]{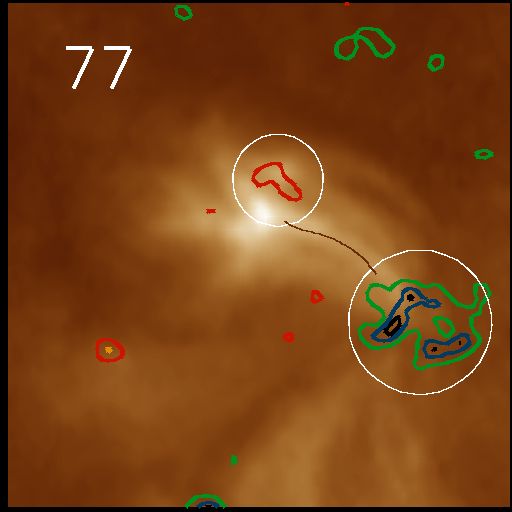}
\includegraphics[scale=0.13]{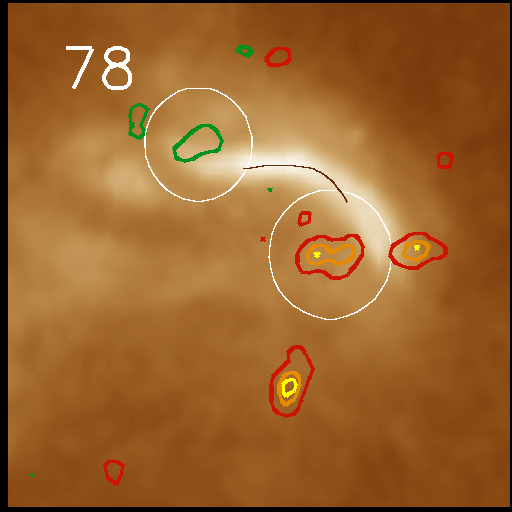}\\
\includegraphics[scale=0.13]{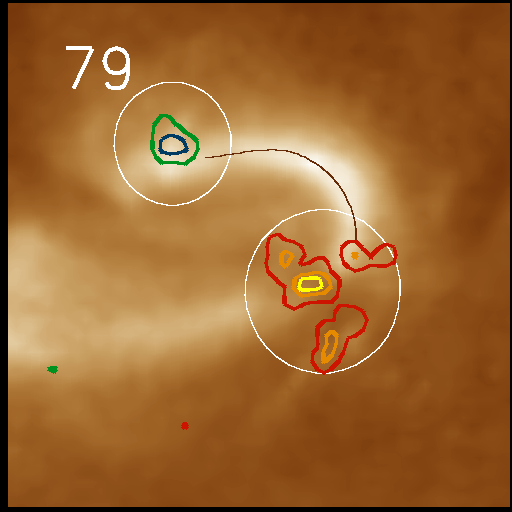}
\includegraphics[scale=0.13]{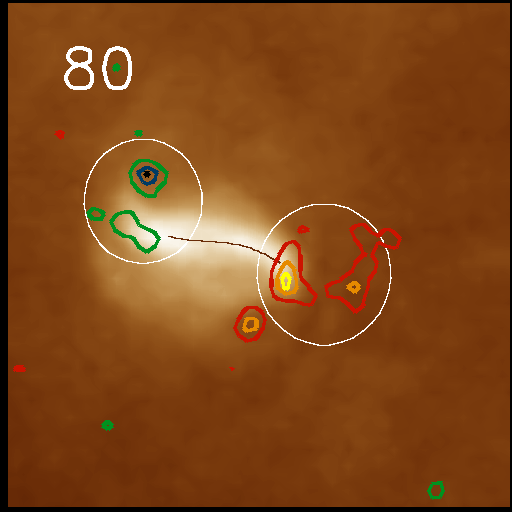}
\includegraphics[scale=0.13]{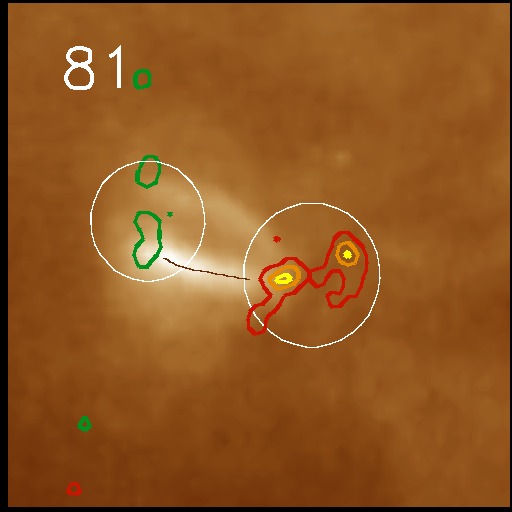}
\includegraphics[scale=0.13]{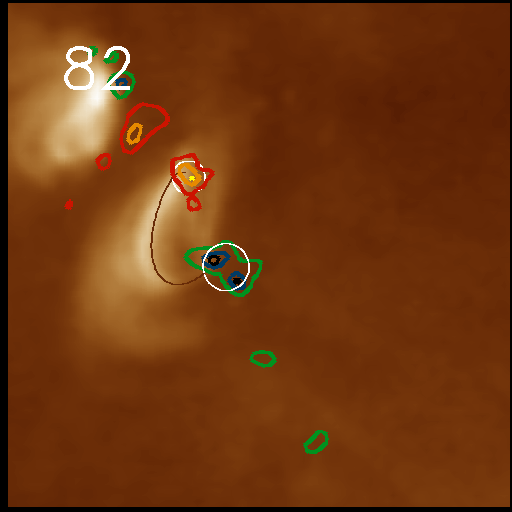}
\includegraphics[scale=0.13]{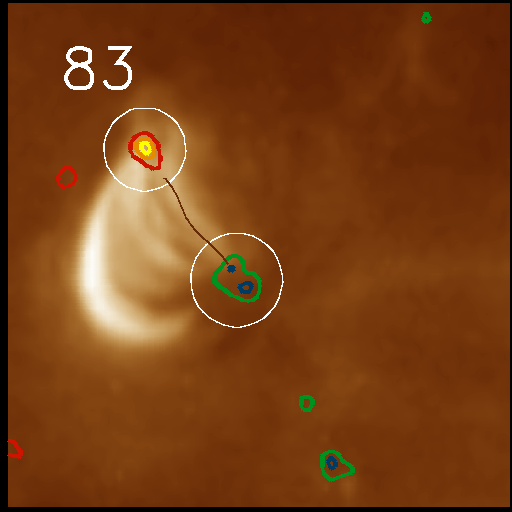}
\includegraphics[scale=0.13]{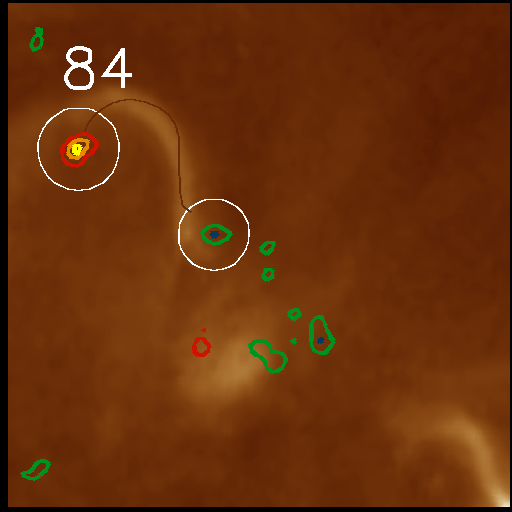}\\
\caption{Same as Fig.~\ref{app-fig1} for loops 43--84.}
\label{app-fig2}
\end{figure*}   
    
\pagebreak
\clearpage  
\begin{figure*}
\vspace{3cm}
\centering
\includegraphics[scale=0.13]{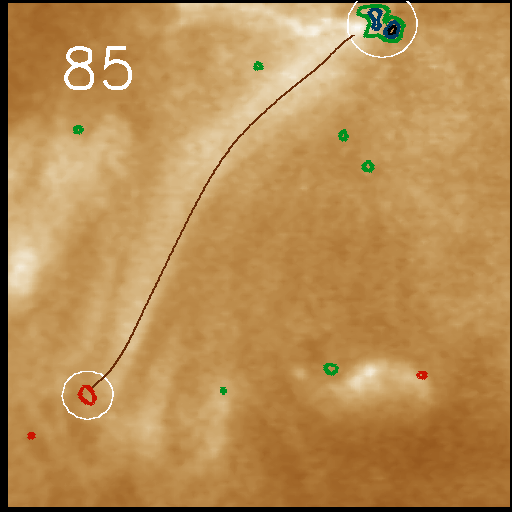}
\includegraphics[scale=0.13]{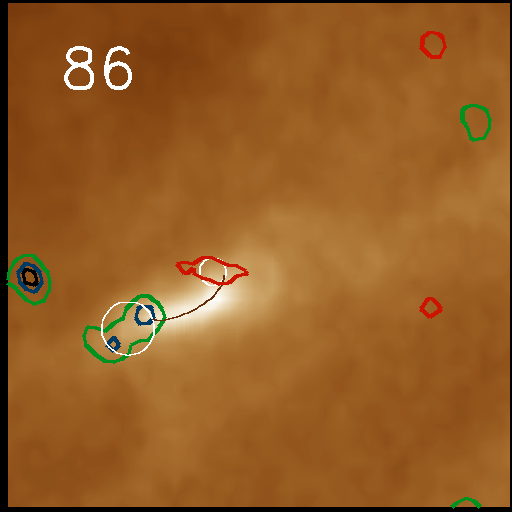}
\includegraphics[scale=0.13]{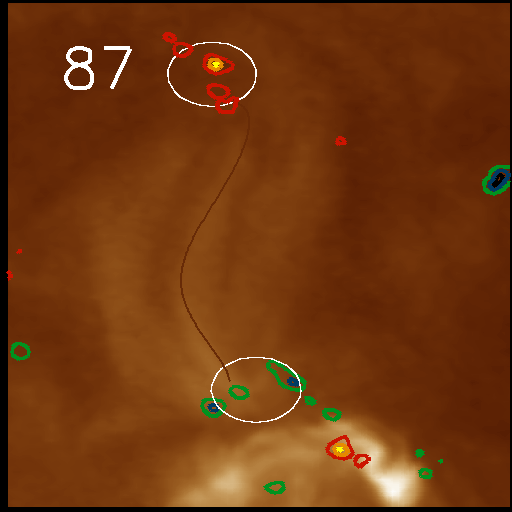}
\includegraphics[scale=0.13]{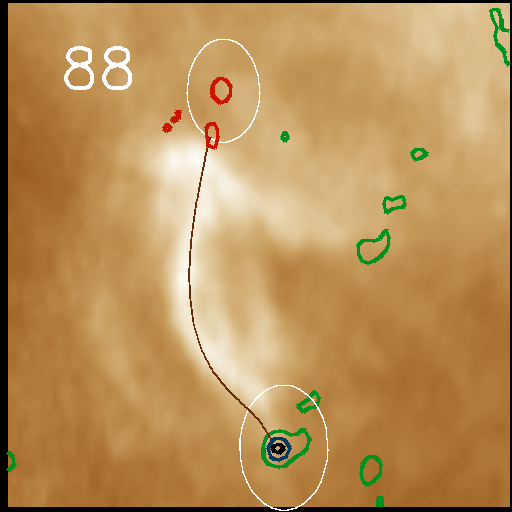}
\includegraphics[scale=0.13]{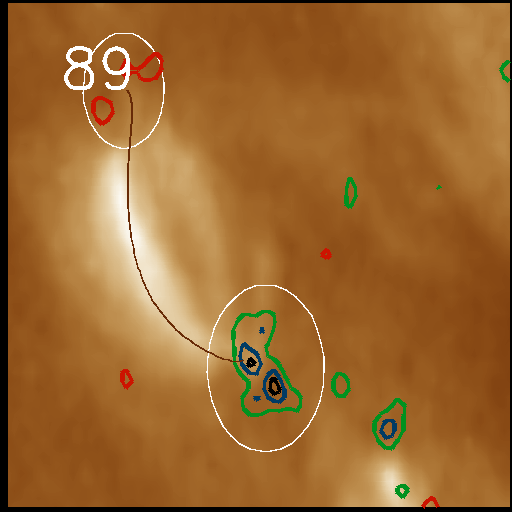}
\includegraphics[scale=0.13]{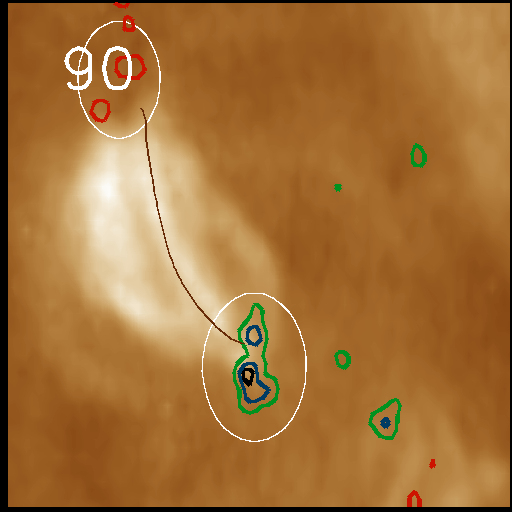}\\
\includegraphics[scale=0.13]{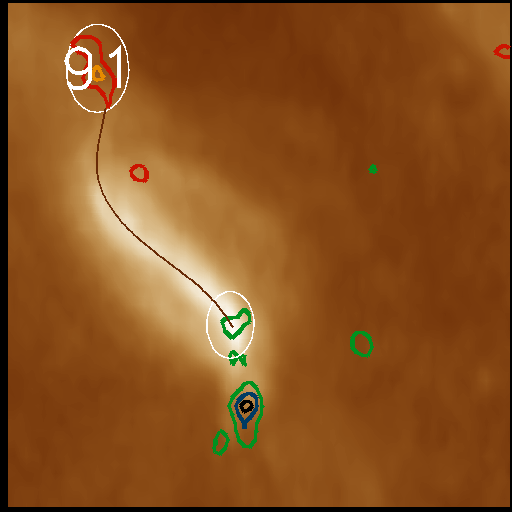}
\includegraphics[scale=0.13]{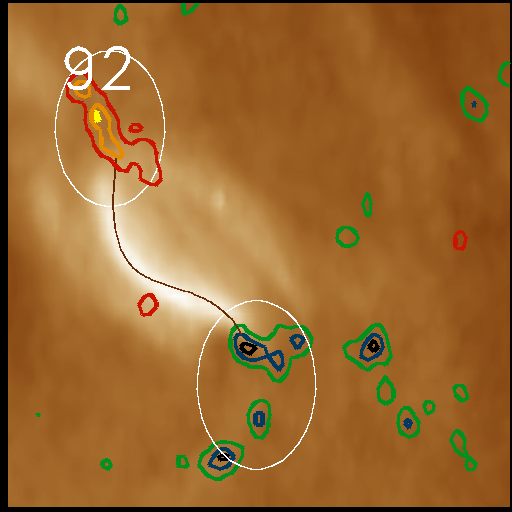}
\includegraphics[scale=0.13]{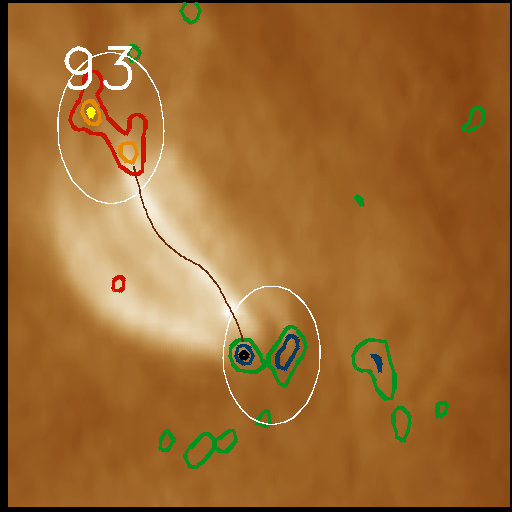}
\includegraphics[scale=0.13]{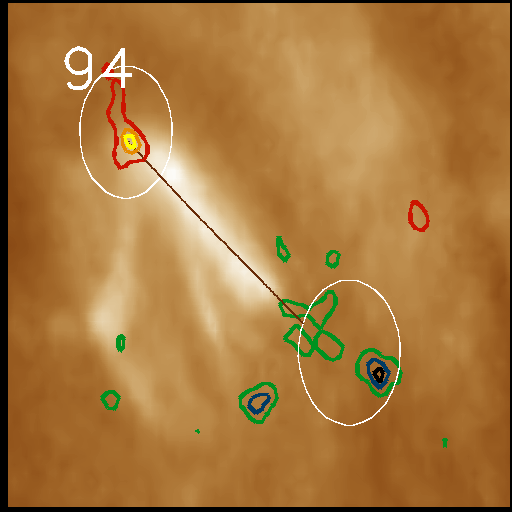}
\includegraphics[scale=0.13]{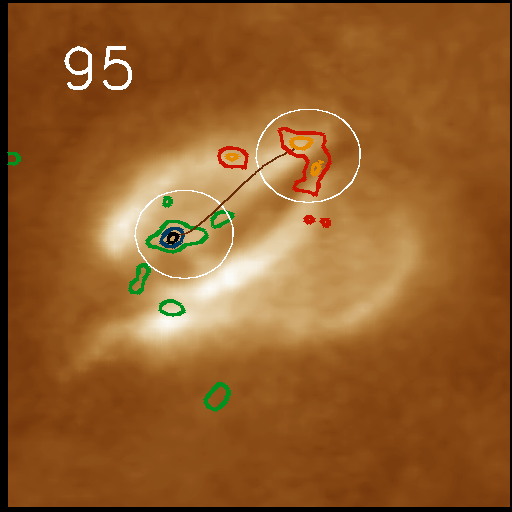}
\includegraphics[scale=0.13]{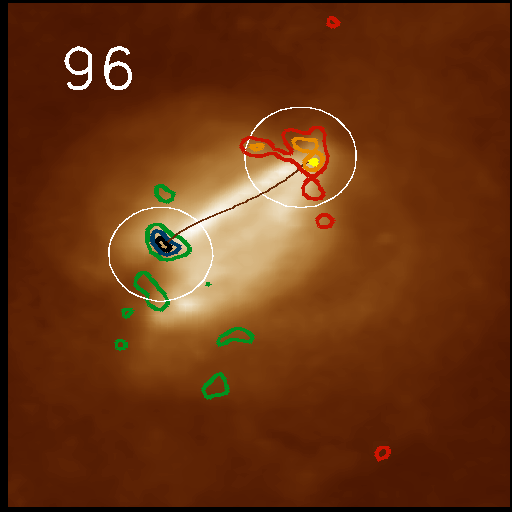}\\
\includegraphics[scale=0.13]{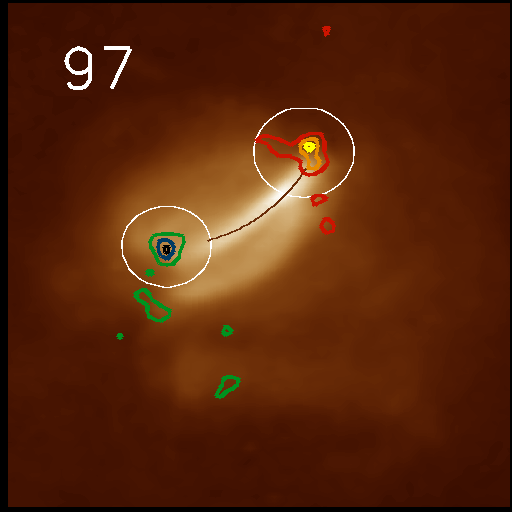}
\includegraphics[scale=0.13]{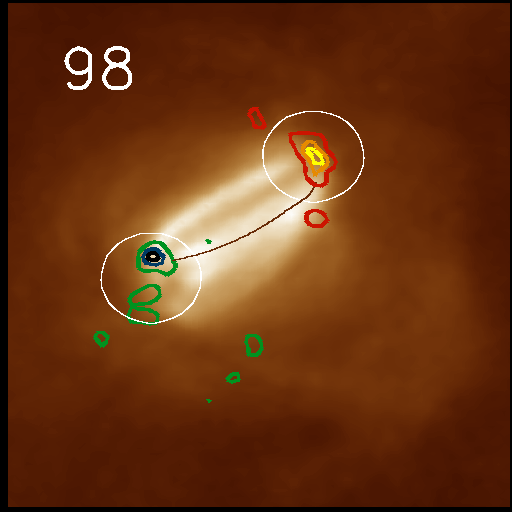}
\includegraphics[scale=0.13]{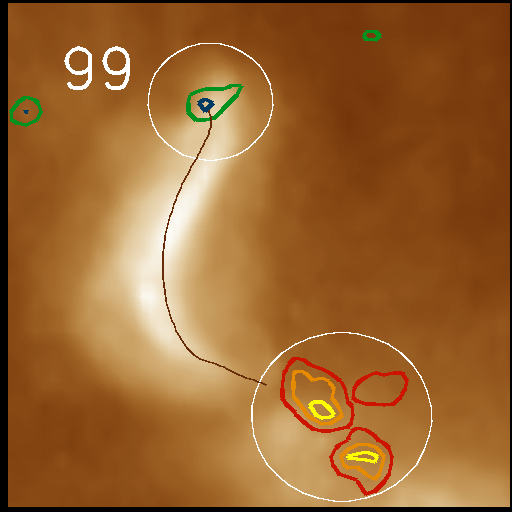}
\includegraphics[scale=0.13]{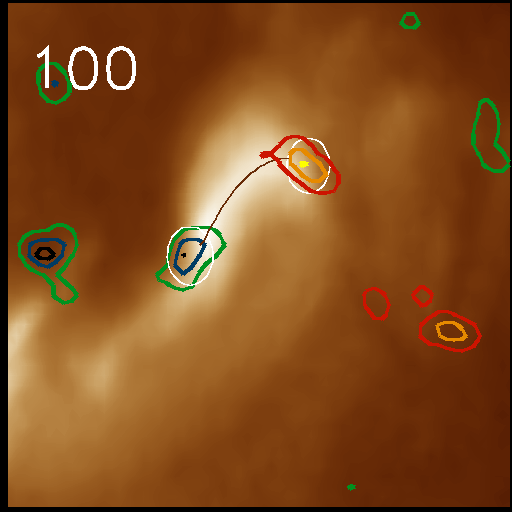}
\includegraphics[scale=0.13]{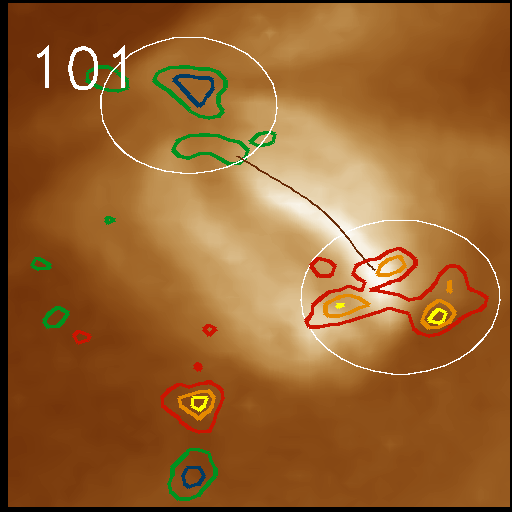}
\includegraphics[scale=0.13]{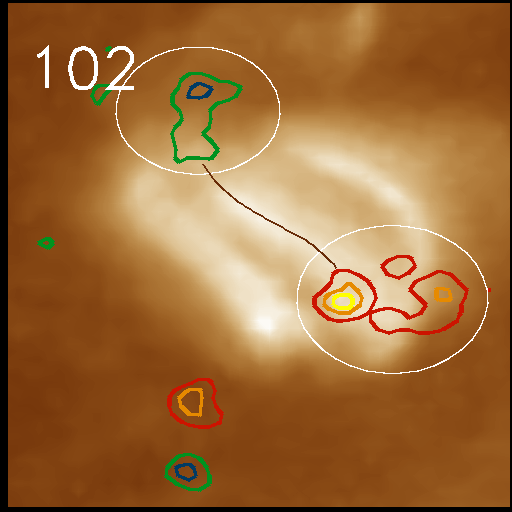}\\ 
\includegraphics[scale=0.13]{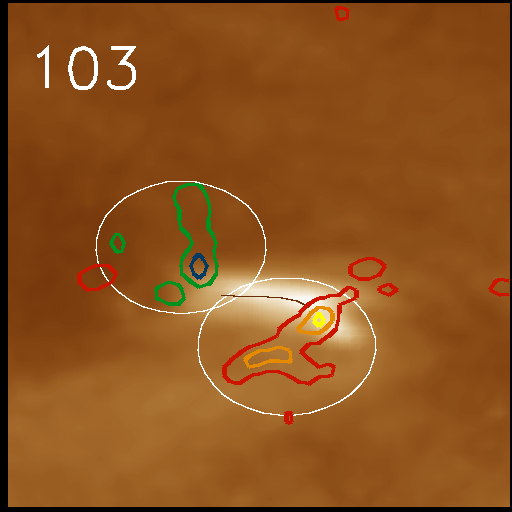}
\includegraphics[scale=0.13]{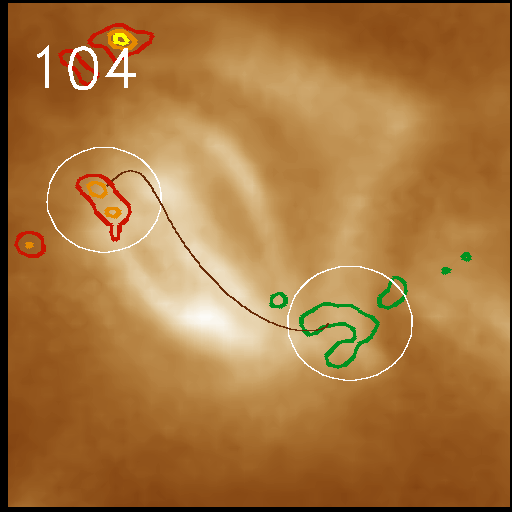}
\includegraphics[scale=0.13]{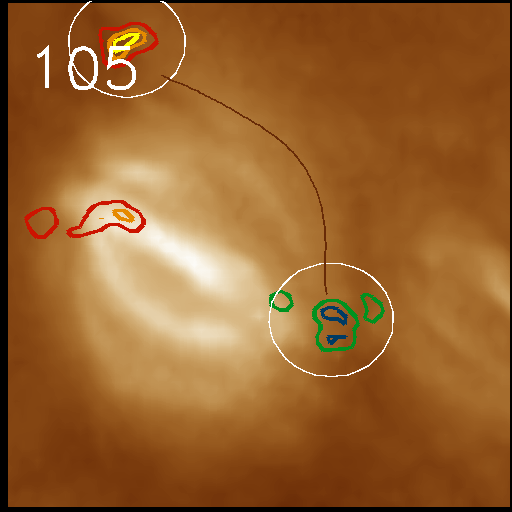}
\includegraphics[scale=0.13]{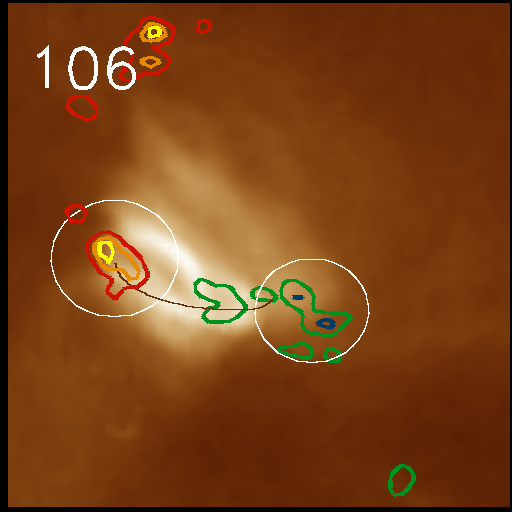}
\includegraphics[scale=0.13]{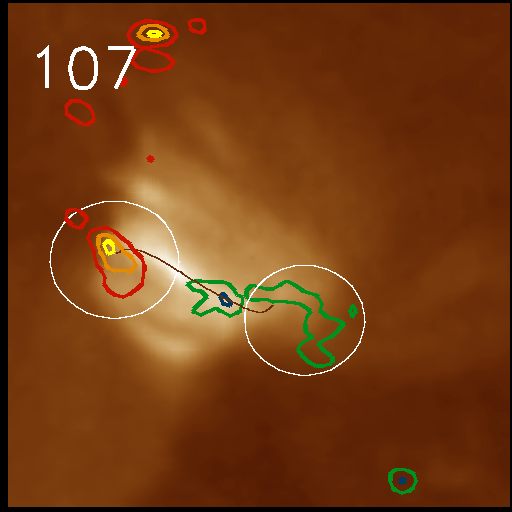}
\includegraphics[scale=0.13]{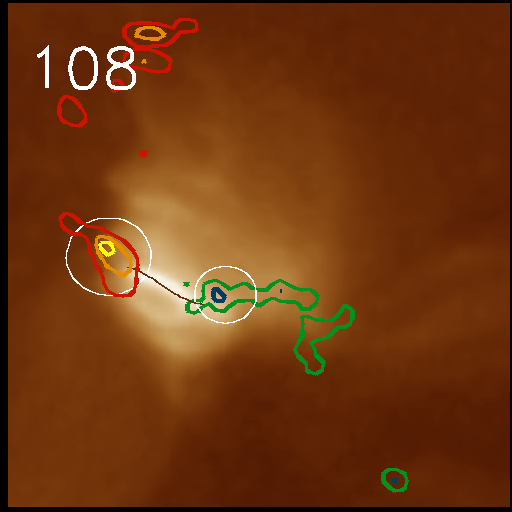}\\ 
\includegraphics[scale=0.13]{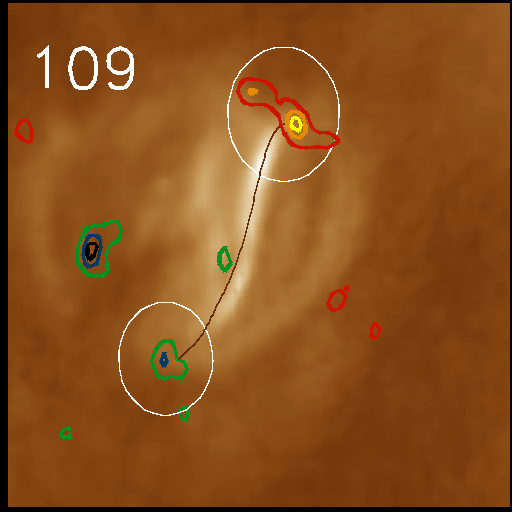}
\includegraphics[scale=0.13]{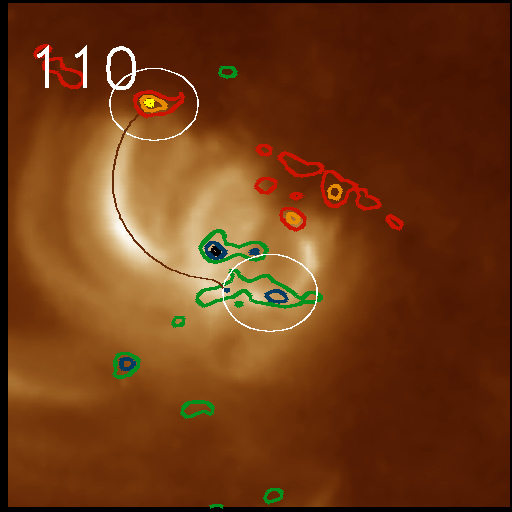}
\includegraphics[scale=0.13]{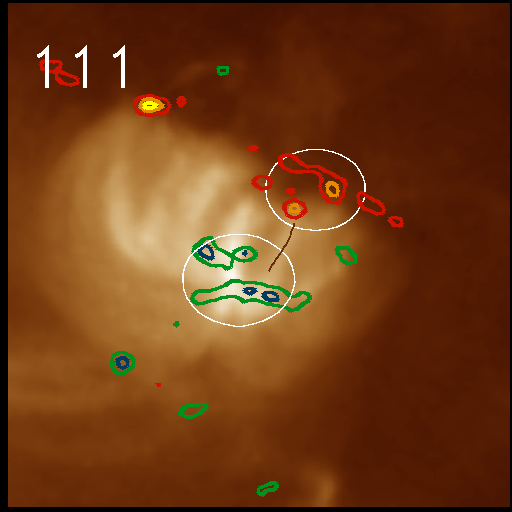}
\includegraphics[scale=0.13]{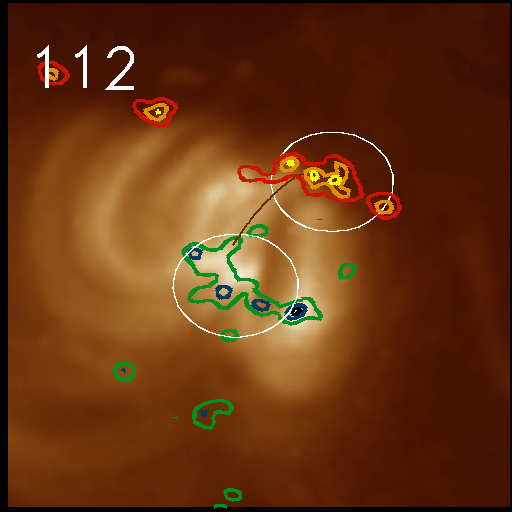}
\includegraphics[scale=0.13]{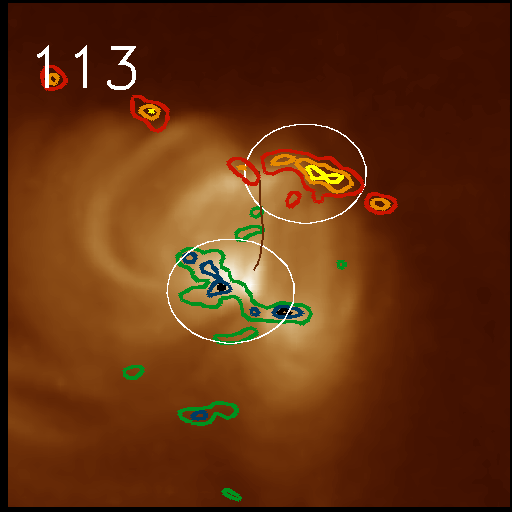}
\includegraphics[scale=0.13]{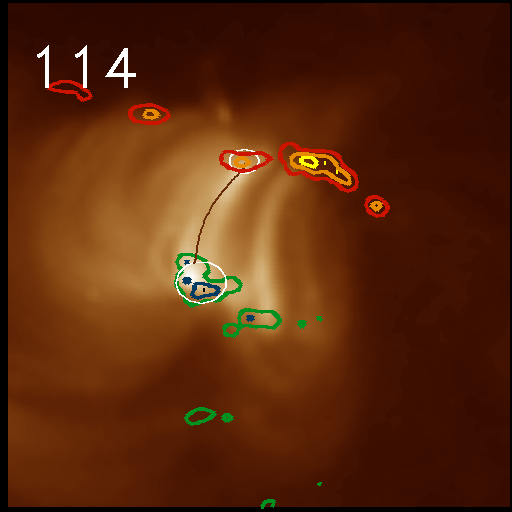}\\ 
\includegraphics[scale=0.13]{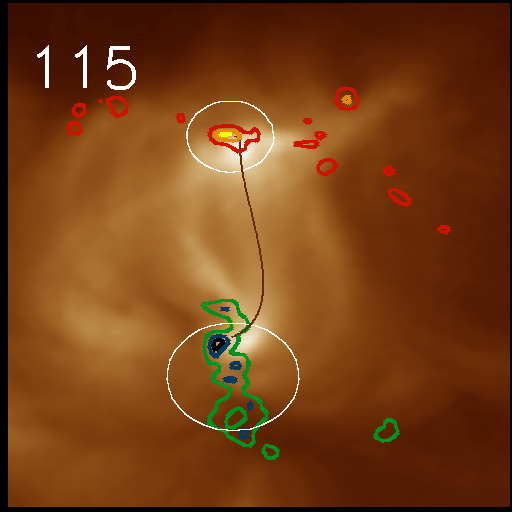}
\includegraphics[scale=0.13]{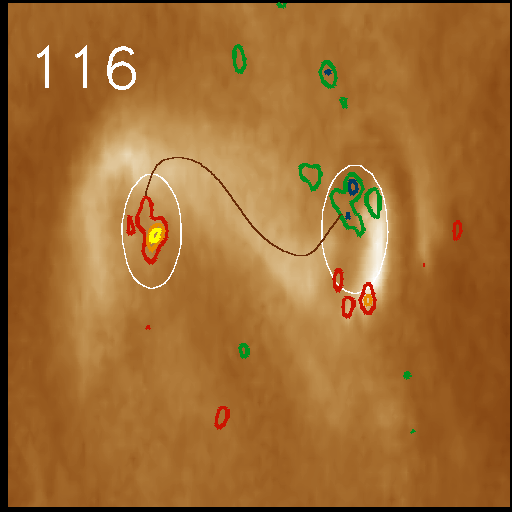}
\includegraphics[scale=0.13]{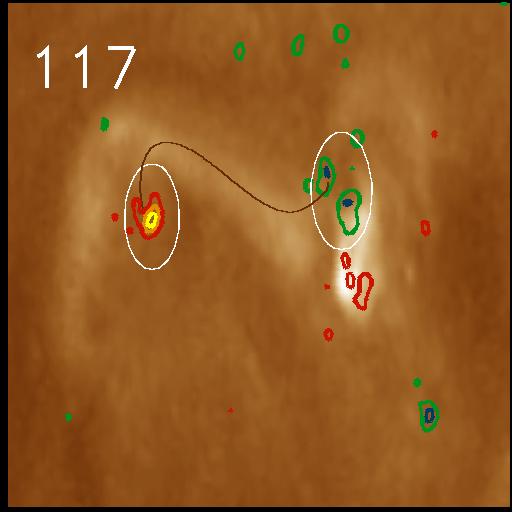}
\includegraphics[scale=0.13]{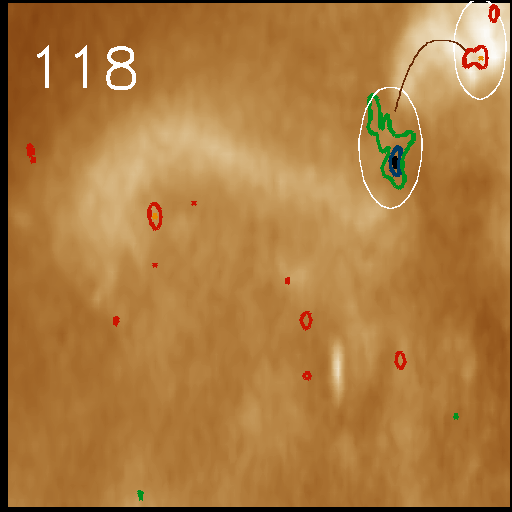}
\includegraphics[scale=0.13]{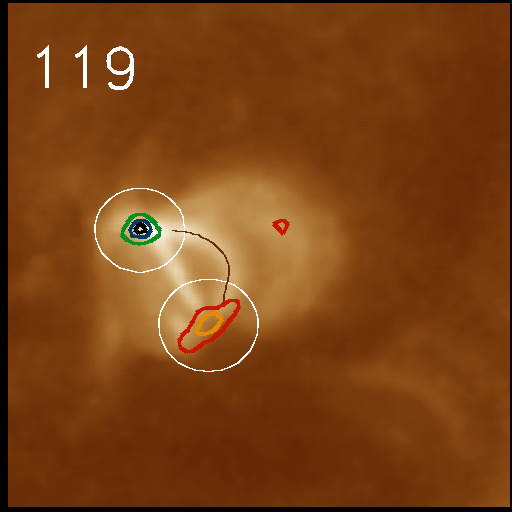}
\includegraphics[scale=0.13]{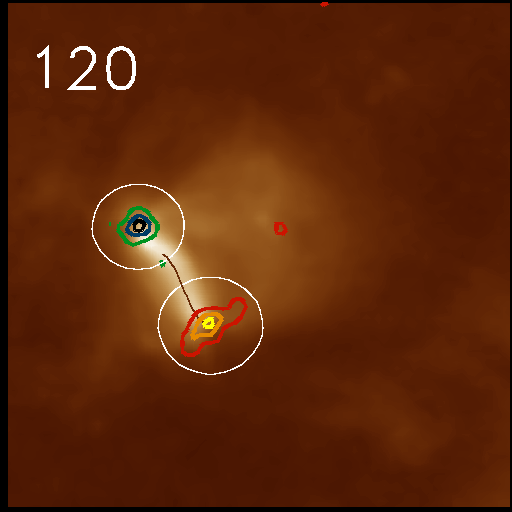}\\
\includegraphics[scale=0.13]{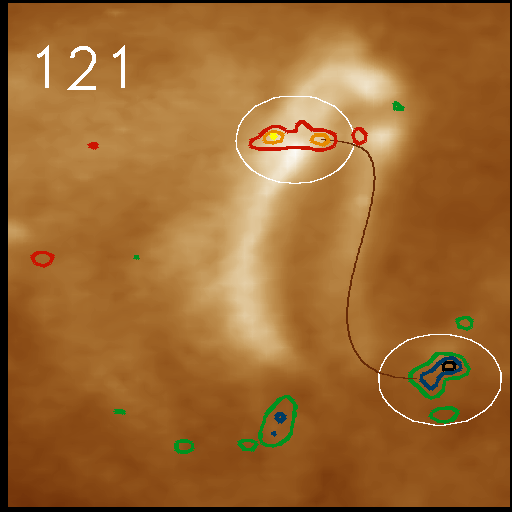}
\includegraphics[scale=0.13]{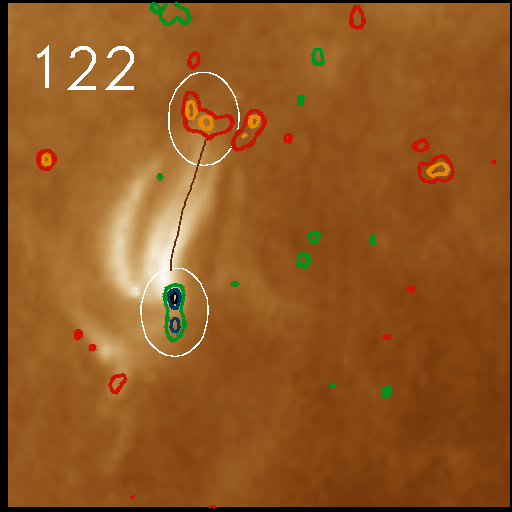}
\includegraphics[scale=0.13]{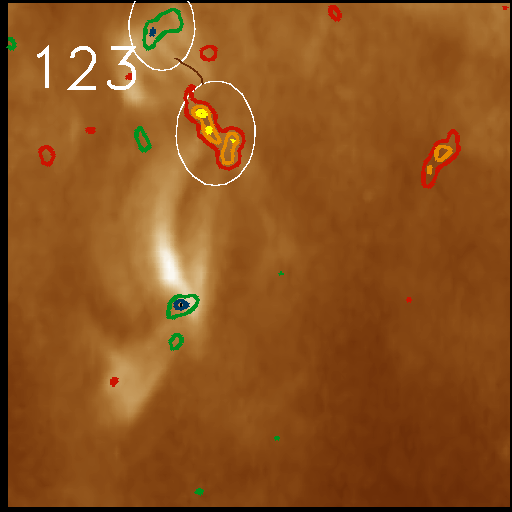}
\includegraphics[scale=0.13]{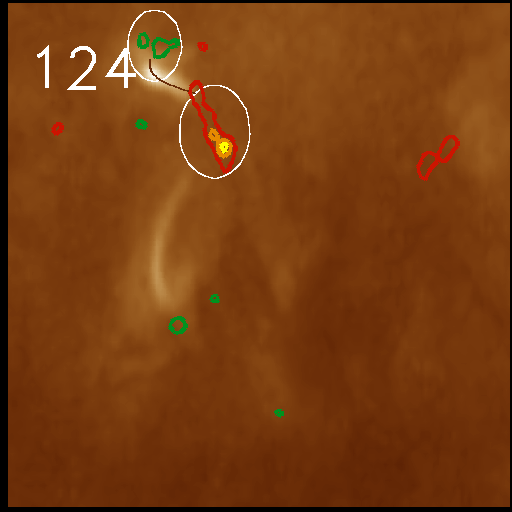}
\includegraphics[scale=0.13]{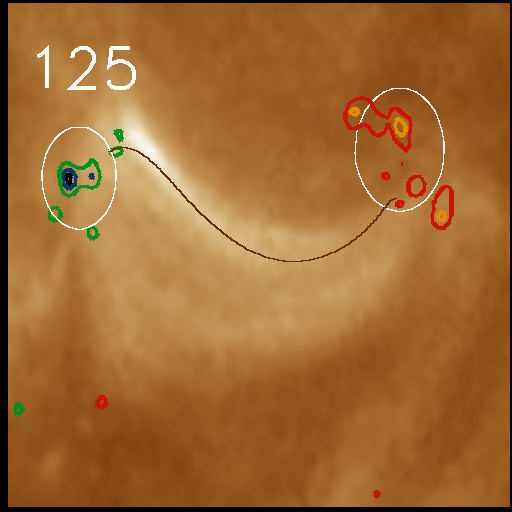}
\includegraphics[scale=0.13]{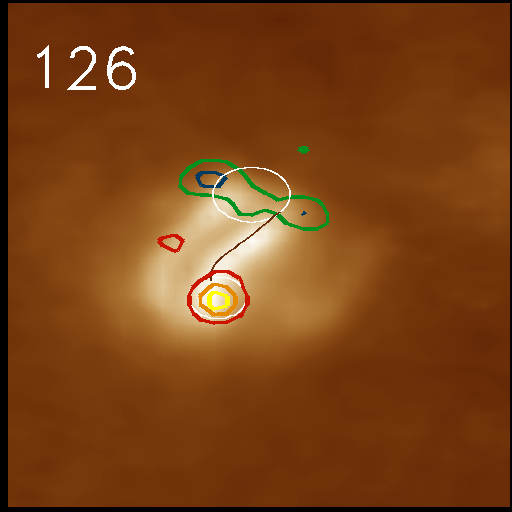}\\
\caption{Same as Fig.~\ref{app-fig1} for loops 85--126.}
\label{app-fig3}
\end{figure*}

 \begin{figure*}
\vspace{3cm}
\centering  
\includegraphics[scale=0.33]{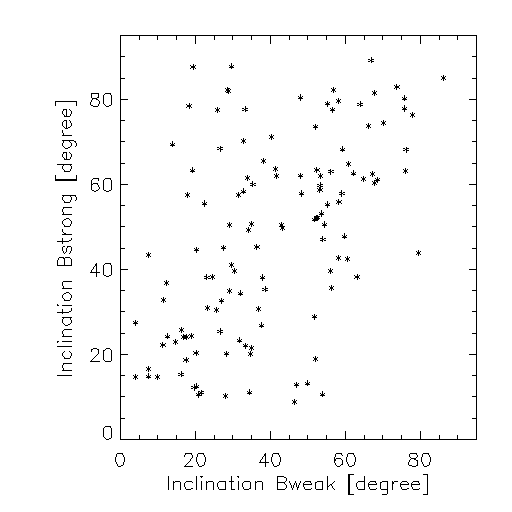}
\includegraphics[scale=0.33]{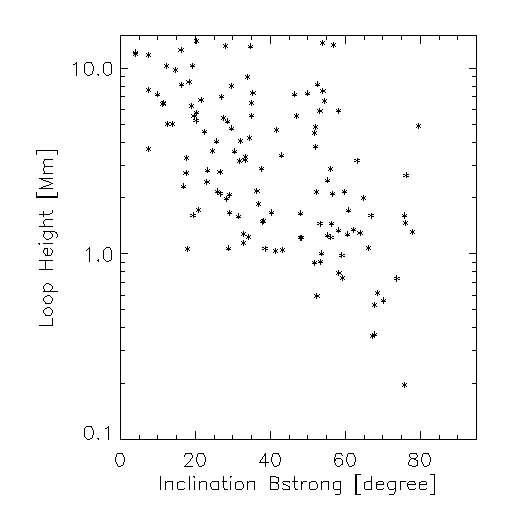}
\includegraphics[scale=0.33]{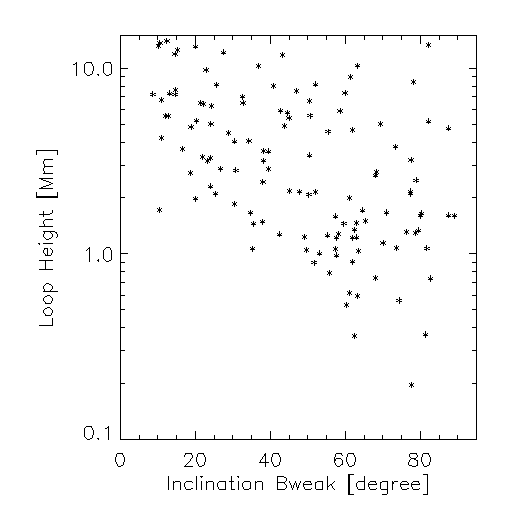}
\caption{Scatter plots of magnetic field line inclination relationships. Left: Inclination at the weak $B$ footpoint vs. the strong. Middle: Inclination at the strong $B$ footpoint vs. loop height. Right: Inclination at the weak $B$ footpoint vs. loop height.
Linear fits: $y=a+b x$ with $a=23.56$ and b=$0.60$ for
 inclination at strong vs. inclination at weak footpoint.
$a=7.01$ and $b=-0.08$ for inclination at strong
footpoint vs. loop height.
$a=7.07$ and b=$-0.07$ for inclination at weak footpoint vs. loop height.}
\label{app-fig4}
\end{figure*}
\end{appendix}

\end{document}